\documentclass[12pt,twoside]{article}

\pdfoutput=1

\usepackage[usenames]{color}
\usepackage{lscape}
\usepackage{amssymb}
\usepackage{rotating}
\usepackage[T1]{fontenc}
\usepackage{graphicx}
\usepackage{fancyhdr}
\usepackage{amsmath}
\usepackage{ae}
\usepackage{aecompl}
\usepackage{hyperref}

\def\pa{\partial}
\def\nn{\nonumber \\}
\def\ov{\overline}

\def\sgn{{\rm sgn}}

\def\hathh{\hat{h}\hat{h}}
\def\hatHH{\hat{H}\hat{H}}
\def\hatss{\hat{s}\hat{s}}
\def\hathH{\hat{h}\hat{H}}
\def\haths{\hat{h}\hat{s}}
\def\hatHs{\hat{H}\hat{s}}

\newlength{\dinwidth}
\newlength{\dinmargin}
\begin{document}

\thispagestyle{empty}

\vspace*{1cm}

\centerline{\Large\bf Blind spots for neutralino dark matter in the NMSSM}

\vspace*{5mm}

\vspace*{5mm} \noindent
\vskip 0.5cm
\centerline{\bf
Marcin Badziak\footnote[1]{mbadziak@fuw.edu.pl},
Marek Olechowski\footnote[2]{Marek.Olechowski@fuw.edu.pl},
Pawe\l\ Szczerbiak\footnote[3]{Pawel.Szczerbiak@fuw.edu.pl}
}
\vskip 5mm

\centerline{\em Institute of Theoretical Physics, Faculty of Physics,
University of Warsaw} 
\centerline{\em ul.\ Pasteura 5, PL--02--093 Warsaw, Poland}

\vskip 1cm

\centerline{\bf Abstract}

\vskip 3mm

Spin-independent cross-section for neutralino dark matter scattering off 
nuclei is investigated in the NMSSM. Several classes of blind spots for 
direct detection of singlino-Higgsino dark matter are analytically identified, 
including such that have no analog in the MSSM. It is shown that mixing  
of the Higgs doublets with the scalar singlet has a big impact on the position 
of blind spots in the parameter space. In particular, this mixing allows 
for more freedom in the sign assignment for the parameters entering the 
neutralino mass matrix, required for a blind spot to occur, as compared 
to the MSSM or the NMSSM with decoupled singlet. Moreover, blind spots may 
occur for any composition of a singlino-Higgsino LSP. Particular attention 
is paid to cases with the singlet-dominated scalar lighter than the 125 GeV Higgs 
for which a vanishing tree-level spin-independent scattering cross-section 
may result from destructive interference between the Higgs and the 
singlet-dominated scalar exchange. Correlations of the spin-independent 
scattering cross-section with the Higgs observables are also discussed.

\newpage

\section{Introduction}

After the recent discovery of the Higgs boson 
\cite{Atlas_discovery,CMS_discovery}, probably the most wanted new particle 
is the one responsible for the observed dark matter (DM) in the Universe. 
Among extensions of the Standard Model (SM) that provide a candidate for 
a dark matter particle, supersymmetric models are most attractive. 
One of the main reasons that kept particle physics community interested 
in supersymmetric models for more than three decades is their ability to 
solve the hierarchy problem of the SM. Moreover, in the simplest 
supersymmetric extensions of the SM the lightest supersymmetric 
particle (LSP) is stable and generically neutral making it a good dark 
matter candidate. In most of the supersymmetry breaking schemes the 
LSP is a neutralino.

One of the most promising ways to search for neutralino dark matter is 
through its direct interactions with nuclei. In the last couple of years 
sensitivity of direct dark matter detection experiments improved by several 
orders of magnitude. The best constraints for the spin-independent (SI)
neutralino-nucleon scattering cross-section (for the DM masses above 6 GeV) 
are provided now by the LUX experiment~\cite{LUX}. In consequence, 
significant portions of the neutralino sector parameter space has been 
excluded by LUX. The constraints will become soon even stronger with the 
forthcoming experiments such as XEXON1T~\cite{XENON1T} and LZ~\cite{LZ}. 
Nevertheless, there are points in the parameter space, so-called
blind spots, for which the neutralino LSP spin-independent scattering 
cross-section 
(almost) vanishes at the tree level. In the vicinity of such blind spots 
the neutralino LSP is not only consistent with the LUX constraints but,  
due to the irreducible neutrino background~\cite{NeutrinoB}, 
might be never detected in direct detection experiments sensitive 
only to the SI scattering cross-section.   
When comparing with the results of DM detection experiments we assume 
that the considered particle is the main component of DM with the relic 
density obtained by the Planck satellite \cite{Planck} (otherwise the 
experimental bounds on the cross-sections should be re-scaled by the 
ratio $\Omega_{\rm observed}/\Omega_{\rm LSP}$).

Conditions for the existence of blind spots have been already identified 
in the Minimal Supersymmetric Standard Model (MSSM). In Ref.~\cite{Hall} 
the conditions for MSSM parameters leading to a vanishing 
Higgs-neutralino-neutralino coupling were found in the limit of decoupled 
heavy Higgs doublet. Additional blind spots in the MSSM originating from 
destructive interference between contributions to the scattering amplitude 
mediated by the 125 GeV Higgs and the heavy Higgs doublet were found in
Ref.~\cite{Wagner}. 
However, the measured Higgs scalar mass strongly motivates extensions of the 
MSSM because the 125 GeV Higgs implies in the MSSM relatively heavy stops 
threatening naturalness of supersymmetry. Substantially lighter stops than 
in the MSSM can be consistent with the 125 GeV Higgs in the Next-to-Minimal 
Supersymmetric Standard Model (NMSSM) \cite{reviewEllwanger} which is the MSSM
supplemented by a gauge singlet chiral superfield. The neutralino sector 
of the NMSSM is richer than that of the MSSM because it contains, 
in addition, the fermionic component of the singlet superfield -- the singlino. 
In some part of the parameter space the LSP has a non-negligible singlino 
component and can be a good dark matter candidate 
\cite{Greene:1986th,Flores:1990bt,Belanger:2005kh} but with different 
properties than those of the LSP in the MSSM. There have been many studies 
of neutralino dark matter in the NMSSM including predictions for its direct 
detection, see 
e.g.~refs.~\cite{NMSSM_DD,NMSSM_DD_Barger,Das:2010ww,Kozaczuk:2013spa,Cao:2013mqa,Han:2014nba} 
and references therein.\footnote{Prospects for indirect detection of Higgsino-singlino LSP have also been studied \cite{Enberg:2015qwa}.} However,
conditions for blind spots in the NMSSM 
have not been discussed in the literature so far.

The main aim of this paper is to investigate conditions for 
SI scattering cross-section blind spots 
for a singlino-Higgsino LSP in the NMSSM. We find a general formula for the 
blind spot condition and study it in the most interesting and 
phenomenologically relevant limiting cases, focusing both on small and 
large $\tan\beta$ regions. First of all, we identify blind spots analogous 
to those for a gaugino-Higgsino LSP in the MSSM originating from a vanishing 
Higgs-neutralino-neutralino coupling \cite{Hall}. Such blind spots
were also found in a general singlet-doublet DM model which mimics NMSSM 
with a Higgsino-singlino DM with a decoupled scalar singlet and heavy 
MSSM-like doublet \cite{mixedDM_Cheung} (see also Ref.~\cite{mixedDM_Calibbi} 
for a recent analysis). However, in our analysis we include also the effects 
of mixing among scalars. We find that inclusion of the mixing 
with the singlet introduces qualitatively new features to the conditions 
for blind spots, e.g.\ allowing certain signs of some parameters that would 
be forbidden if such mixing is neglected. Secondly, we find blind 
spots analogous to those in the MSSM with the effect of the heavy doublet
taken into account \cite{Wagner} and generalize them to the case with 
the Higgs-singlet mixing included.

Finally, we investigate in great detail the region of the NMSSM parameter 
space with the singlet-dominated scalar lighter than 125 GeV, which is 
entirely new with respect to the MSSM. This region is particularly interesting 
because the Higgs-singlet mixing can increase the Higgs boson mass by up to 
about 6 GeV \cite{BaOlPo}. While this enhancement of the Higgs mass by mixing 
effects can be present both for small and large $\tan\beta$, it is worth 
emphasizing that for large (or moderate) values of $\tan\beta$ this is a unique 
way to have lighter stops than in the MSSM. Moreover, for large $\tan\beta$ the 
singlet-dominated scalar coupling to bottom quarks can be strongly suppressed 
relaxing the LEP constraints on scalars and allowing a substantial correction 
to the Higgs mass from mixing for a wide range of singlet masses 
between about 60 and 110 GeV \cite{BaOlPo} (for small $\tan\beta$ a sizable 
correction from mixing is allowed only for the singlet mass in the 
vicinity of the LEP excess at 98 GeV \cite{LEP2}).

In the case of a light singlet-dominated scalar with sizable mixing with the 
Higgs scalar, the SI scattering cross-section is generically large, even for 
not too large values of $\lambda$. The main reason for this is that such 
a singlet-dominated scalar also mediates the SI scattering cross-section and 
the corresponding amplitude may even dominate over the one with the 
SM-like Higgs boson exchange due to the enhancement by a small mass of the 
singlet-dominated scalar. This phenomenon was identified long before the Higgs 
scalar discovery \cite{NMSSM_DD}. 
Recently, points in the parameter space of the NMSSM with strongly suppressed 
SI direct detection cross-section, consistent with LUX constraints and in some 
cases even below the irreducible neutrino background for direct detection 
experiments, were found using sophisticated numerical scans of 
semi-constrained NMSSM \cite{NMSSM_DD_EllwangerHugonie}. 
However, in Ref.~\cite{NMSSM_DD_EllwangerHugonie} no explanation was
given why such points exist and what are the conditions for the NMSSM 
parameters required for this suppression to occur. In the present paper 
we provide analytic understanding for the existence of blind spots 
in the NMSSM with light singlet-dominated scalar and a Higgsino-singlino LSP. 
Such blind spots follow from a destructive interference between the singlet 
and Higgs exchange in the scattering amplitude. We also discuss the influence 
of a strongly suppressed coupling of the singlet-dominated scalar to $b$ quarks 
which is important at large $\tan\beta$. In particular, we find that the 
presence of a light singlet-dominated scalar gives much more freedom in the 
LSP composition and, especially for a singlino-dominated LSP, in sign 
assignments of various NMSSM parameters required for obtaining a blind spot.

The rest of the paper is organized as follows. In section \ref{NMSSM} we review 
some features of the Higgs and neutralino sector of the NMSSM that are 
important for the analysis of blind spots. In section \ref{sec:cross-section} 
SI scattering cross-section in the NMSSM is reviewed and general formulae for 
neutralino blind spots are derived. In the remaining sections blind spot 
conditions are analyzed in detail in several physically interesting
cases and approximations. In section \ref{sec:fh} only SM-like Higgs scalar 
exchange is taken into account. In section \ref{ref:sec_fhH} the interference 
effects between two doublet-dominated scalars are analyzed, while section 
\ref{ref:sec_fhs} is focused on the case with a light singlet-dominated scalar 
in which interference effects between such light scalar and the SM-like 
Higgs scalar become important. 
Our main findings are summarized in section \ref{sec:sum}.

\section{Higgs and neutralino sector of the NMSSM}
\label{NMSSM}

Several versions of NMSSM has been proposed so far \cite{reviewEllwanger}. 
We would like to keep our discussion as general as possible 
so we assume that the NMSSM specific part of the superpotential 
and the soft terms have the following general forms:
\begin{equation}
\label{W}
W_{\rm NMSSM}= \left(\mu_{H_uH_d}+\lambda S\right)H_uH_d + f(S) \,,
\end{equation}
\begin{align}
-{\cal{L}}_{\rm soft}
\supset
&\,\,
m_{H_u}^2\left|H_u\right|^2+m_{H_d}^2\left|H_d\right|^2+m_{S}^2\left|S\right|^2
\nn
&\label{Lsoft}
+\left(
A_\lambda\lambda H_u H_d S +\frac13A_\kappa\kappa S^3
+m_3^2H_uH_d + \frac12 m_S'^2S^2 + \xi_SS
+{\rm h.c.}\right)\,,
\end{align}
where $S$ is an additional SM-singlet superfield. The first term in (\ref{W}) 
is the source of the effective Higgsino mass parameter, 
$\mu_{\rm eff}\equiv\mu_{H_uH_d}+\lambda v_s$ 
(we drop the subscript ``eff'' in the rest of the paper). 
Using the shift symmetry of $S$ we can put $\mu_{H_uH_d}=0$.
In the simplest version, known as the scale-invariant NMSSM, 
$m_3^2=m_S'^2=\xi_S=0$ while $f(S)\equiv\kappa S^3/3$. In more general 
models $f(S)\equiv\xi_F S+\mu'S^2/2+\kappa S^3/3$.

There are three neutral CP-even scalar fields, 
$H_u$, $H_d$, $S$ which are the real parts of excitations 
around the real vevs, $v_u\equiv v \sin\beta$, $v_d\equiv v \cos\beta$, 
$v_s$ with $v^2=v_u^2 + v_d^2\approx (174 {\rm GeV})^2$,  
of the neutral components of the doublets $H_u$, $H_d$ and the singlet $S$ 
(we use the same notation for the doublets and the singlet as for 
the real parts of their neutral components). 
It is more convenient for us to work in the basis 
$\left(\hat{h}, \hat{H}, \hat{s}\right)$, where 
$\hat{h}=H_d\cos\beta + H_u\sin\beta$, 
$\hat{H}=H_d\sin\beta - H_u\cos\beta$ and $\hat{s}=S$. The $\hat{h}$ field 
has exactly the same couplings to the gauge bosons and fermions as the SM 
Higgs field. In this basis the scalar mass squared matrix reads:
\begin{equation}
\label{tilde_M^2}
 {M}^2=
\left(
\begin{array}{ccc}
  {M}^2_{\hathh} & {M}^2_{\hathH} & {M}^2_{\haths} \\[4pt]
   {M}^2_{\hathH} & {M}^2_{\hatHH} & {M}^2_{\hatHs} \\[4pt]
   {M}^2_{\haths} & {M}^2_{\hatHs} & {M}^2_{\hatss} \\
\end{array}
\right) \,,
\end{equation}
where
\begin{align}
\label{Mhh}
 &{M}^2_{\hathh} = M_Z^2\cos^2\left(2\beta\right)
+ \lambda^2 v^2\sin^2\left(2\beta\right) 
+(\delta m_h^2)^{\rm rad}, \\
\label{MHH}
&{M}^2_{\hatHH} = (M_Z^2-\lambda^2 v^2)\sin^2\left(2\beta\right) 
+ 
\frac{2}{\sin\left(2\beta\right)}
\left(\mu A_{\lambda}+ \frac{\mu\langle\pa_S f\rangle}{v_s}
+m_3^2\right), 
\\
\label{Mss}
&{M}^2_{\hatss} =  
\frac12\lambda v^2 \sin2\beta
\left(\frac{\Lambda}{v_s}-\left<\partial^3_Sf\right>\right)
+\langle(\partial_S^2f)^2 + \partial_S f\,\partial_S^3 f\rangle
-\frac{\langle\partial_S f\,\partial_S^2 f\rangle}{v_s}
+A_\kappa\kappa v_s -\frac{\xi_S}{v_s}\,, \\
\label{MhH}
& {M}^2_{\hathH} = \frac{1}{2}(M^2_Z-\lambda^2 v^2)\sin4\beta, \\
\label{Mhs}
& {M}^2_{\haths} =  \lambda v (2\mu-\Lambda \sin2\beta), \\
\label{MHs}
& {M}^2_{\hatHs} = \lambda v \Lambda \cos2\beta ,
\end{align}
and $\Lambda \equiv A_{\lambda}+\langle\pa_S^2 f\rangle$. 
We neglected all the radiative corrections except those 
to ${M}^2_{\hathh}$ which we parametrize by $(\delta m_h^2)^{\rm rad}$. 
The mass eigenstates of ${M}^2$, denoted by $h_i$ 
(with $h_i=h,H,s$), 
are expressed in terms of the hatted fields with the help of the 
diagonalization matrix $\tilde{S}$:\footnote
{
The matrix $\tilde{S}$ is related to the commonly used Higgs mixing 
matrix $S$ by a rotation by the angle $\beta$ in the 2-dimensional 
space of the weak doublets. 
}
\begin{equation}
\label{hat-S}
h_i
=\tilde{S}_{h_i\hat{h}}\hat{h}
+\tilde{S}_{h_i\hat{H}}\hat{H}
+\tilde{S}_{h_i\hat{s}}\hat{s}
\,.
\end{equation}
We will refer to the eigenvalue $h$ as the Higgs scalar 
and identify it with the 125 GeV scalar discovered by the LHC 
experiments.

The neutralino mass matrix in NMSSM is 5-dimensional.
However, in this work we assume that gauginos are heavy and thus 
we focus on the sub-matrix describing the three lightest neutralinos:
\begin{equation}
\label{M_chi}
 {M_{\chi^0}}=
\left(
\begin{array}{ccc}
  0 & -\mu & -\lambda v \sin\beta \\[4pt]
  -\mu & 0 & -\lambda v \cos\beta \\[4pt]
  -\lambda v \sin\beta & -\lambda v \cos\beta 
& \langle\partial_S^2f\rangle \\
\end{array}
\right) \,.
\end{equation}
Trading the model dependent term $\langle\partial_S^2f\rangle$
for one of the eigenvalues, $m_{\chi_j}$, of the above neutralino mass
matrix we find the following (exact at the tree level) relations 
for the neutralino diagonalization matrix 
elements:\footnote{
We consider only the $3\times3$ sub-matrix \eqref{M_chi} but we keep 
the notation from the full $5\times5$ neutralino mass matrix i.e.\
$N_{j3}$, $N_{j4}$ and $N_{j5}$ denote, respectively, the two Higgsino 
and the singlino components of the $j$-th neutralino mass eigenstate. 
The mathematical structure of this matrix is very similar to $3\times3$ 
sub-matrix mixing higgsino with one of the gauginos in the MSSM. Many useful 
formulae that follow from this matrix can be found in the Appendix of 
Ref.~\cite{underabundant} with obvious substitutions of the MSSM parameters 
into the NMSSM ones sitting in the corresponding entries of the $3\times3$ 
sub-matrix.
}
\begin{align}
\label{Nj3Nj5}
\frac{N_{j3}}{N_{j5}}
=
\frac{\lambda v}{\mu}
\,
\frac{(m_{\chi_j}/\mu)\sin\beta-\cos\beta}
{1-\left(m_{\chi_j}/\mu\right)^2}
\,,\\[4pt]
\label{Nj4Nj5}\frac{N_{j4}}{N_{j5}}
=
\frac{\lambda v}{\mu}
\,
\frac{(m_{\chi_j}/\mu)\cos\beta-\sin\beta}
{1-\left(m_{\chi_j}/\mu\right)^2}
\,,
\end{align}
where $j=1,2,3$ and $|m_{\chi_1}|\le|m_{\chi_2}|\le|m_{\chi_3}|$.
Later we will be interested mainly in the LSP corresponding to $j=1$, 
so to simplify the notation we will use $m_{\chi}\equiv m_{\chi_1}$. 
Notice that the physical (positive) LSP mass equals to 
$m_{\rm LSP}\equiv|m_{\chi}|$.
The sign of $m_{\chi}$ is the same as that of the diagonal singlino 
entry $\langle\partial_S^2f\rangle$ in the neutralino mass matrix 
\eqref{M_chi}. 
For $|\langle\partial_S^2f\rangle|<|\mu|$ this is obvious. For bigger
values of $|\langle\partial_S^2f\rangle|$ it is also true. In this case 
the two lightest neutralinos are Higgsino-dominated corresponding 
to the mass eigenstates close to $\mu$ and $-\mu$. The lighter of them is the
one which mixes more strongly with the singlino, and generally the mixing is 
stronger between states with the diagonal terms of the same sign 
(unless the corresponding off-diagonal term is exceptionally small).

Using eqs.~\eqref{Nj3Nj5} and \eqref{Nj4Nj5} and the fact that the gauginos 
are decoupled,  
we can express the ratio of the Higgsino to the singlino components 
of the LSP as the following function of the LSP mass and the ratio 
$(\lambda v)/\mu$:
\begin{equation}
\label{Higgsino/singlino}
\frac{1-N_{15}^2}{N_{15}^2}
=
\left(\frac{\lambda v}{\mu}\right)^{\!\!2}
\frac{1+\left(m_\chi/\mu\right)^2-2{(m_\chi}/{\mu})\sin2\beta}
{\left[1-\left({m_\chi}/{\mu}\right)^2\right]^2}
\,.
\end{equation}

In our discussion we will consider only positive values of $\lambda$. 
The results for negative $\lambda$ are exactly the same due to the invariance 
under the transformation $\lambda\to-\lambda$, 
$\kappa\to-\kappa$, $\xi_S\to-\xi_S$, $\xi_F\to-\xi_F$, $S\to-S$ 
with other fields and couplings unchanged. 

\section{Spin-independent scattering cross-section}
\label{sec:cross-section}

The spin-independent cross-section for the LSP interacting 
with the nucleus with the atomic number $Z$ and the mass number $A$ 
is given by
\begin{equation}
\sigma_{\rm SI}
=
\frac{4\mu^2_{\rm red}}{\pi}\,\frac{\left[Zf^{(p)}+(A-Z)f^{(n)}\right]^2}{A^2}
\,,
\end{equation}
where $\mu^2_{\rm red}$ is the reduced mass of the nucleus and the LSP. Usually, 
the experimental limits concern the cross section $\sigma_{\rm SI}$ defined as 
the arithmetic mean of $\sigma_{\rm SI}^{(p)}$ and $\sigma_{\rm SI}^{(n)}$. Thus, 
in the rest of the paper we will follow this convention. When the squarks 
are heavy the effective couplings $f^{(N)}$ ($N=p,n$) 
are dominated by the t-channel exchange of the CP-even scalars
\cite{JuKaGr}:
\begin{equation}
\label{fN}
f^{(N)}
\approx
\sum_{i=1}^3
f^{(N)}_{h_i}
\equiv
\sum_{i=1}^3
\frac{\alpha_{h_i\chi\chi}\alpha_{h_iNN}}{2m_{h_i}^2}
\,.
\end{equation}
The couplings of the $i$-th scalar to the LSP and to the 
nucleon are given, respectively, by
\begin{eqnarray}
\alpha_{h_i\chi\chi}
\!\!&=&\!\!
\sqrt{2}\lambda
\left(S_{i1}N_{14}N_{15}+S_{i2}N_{13}N_{15}+S_{i3}N_{13}N_{14}\right)
-\sqrt{2}\kappa S_{i3}N_{15}^2
\nn
\label{alpha-h00-S}
\!\!&+&\!\!
g_1\left(S_{i1}N_{11}N_{13}-S_{i2}N_{11}N_{14}\right)
-g_2\left(S_{i1}N_{12}N_{13}-S_{i2}N_{12}N_{14}\right)
\end{eqnarray}
and
\begin{equation}
\label{alpha-hNN-S}
\alpha_{h_iNN}
=
\frac{m_N}{\sqrt{2}v} 
\left(
\frac{S_{i1}}{\cos\beta}F^{(N)}_d
+\frac{S_{i2}}{\sin\beta}F^{(N)}_u
\right)\,.
\end{equation}
In the last equation we introduced the combinations 
$F^{(N)}_d=f^{(N)}_d+f^{(N)}_s+\frac{2}{27}f^{(N)}_G$
and $F^{(N)}_u=f^{(N)}_u+\frac{4}{27}f^{(N)}_G$ 
of the form factors $f^{(N)}_q=m_N^{-1}\left<N|m_qq\bar{q}|N\right>$ 
(for $q=u,d,s$) and $f^{(N)}_G=1-\sum_{q=u,d,s}f^{(N)}_q$.
There is still some inconsistency in the literature 
regarding the values of these form factors.  
In our numerical calculations we will take them to be:  
$f^{(p)}_{u} = 0.0153$, $f^{(p)}_{d} = 0.0191$,
$f^{(p)}_{s} = 0.048$, $f^{(p)}_{G} = 0.921$, 
$f^{(n)}_{u} = 0.0107$,
$f^{(n)}_{d} = 0.0273$, $f^{(n)}_{s} = 0.0447$, 
$f^{(n)}_{G} = 0.917$,
which gives the following values of $F$'s:
$F^{(p)}_{u}\approx 0.152$, $F^{(p)}_{d}\approx 0.132$, 
$F^{(n)}_{u}\approx 0.147$, $F^{(n)}_{d}\approx 0.140$~\cite{Belanger:2013oya}.

The couplings of the scalar particles in 
eqs.\ (\ref{alpha-h00-S}) and (\ref{alpha-hNN-S}) 
are expressed in terms of the diagonalization matrices for the 
scalars and neutralinos ($S$ and $N$, respectively) 
written in the usual weak bases. However, for our purposes 
it will be more convenient to use the scalar diagonalization matrix 
$\tilde{S}$ defined in (\ref{hat-S}) for the rotated basis 
($\hat{h}$,$\hat{H}$,$\hat{s}$).
Moreover, we are interested in the situation when the LSP is 
Higgsino-singlino like with negligible contributions from gauginos
 i.e.\ $N_{11}\approx 0\approx N_{12}$. 
Then, the expressions (\ref{alpha-h00-S}) and (\ref{alpha-hNN-S}) 
are approximated by:
\begin{align}
\alpha_{h_i\chi\chi}
\approx
\sqrt{2}\lambda 
&\left[
\tilde{S}_{{h_i}\hat{h}}N_{15}\left(N_{13}\sin\beta+N_{14}\cos\beta\right)
+\tilde{S}_{{h_i}\hat{H}}N_{15}\left(N_{14}\sin\beta-N_{13}\cos\beta\right)
\right.\nn&\,\,\,\,\left.
+\tilde{S}_{{h_i}\hat{s}}
\left({N_{13}}{N_{14}}-\frac{\kappa}{\lambda}N_{15}^2\right)
\right]
\,,
\label{alpha-h00_G}
\end{align}

\begin{equation}
\label{alpha-hNN_G}
\alpha_{h_iNN}
\approx
\frac{m_N}{\sqrt{2}v} 
\left[
\tilde{S}_{{h_i}\hat{h}}\left(F^{(N)}_d+F^{(N)}_u\right)
+\tilde{S}_{{h_i}\hat{H}} \left(\tan\beta F^{(N)}_d-\frac{1}{\tan\beta} F^{(N)}_u \right)
\right]\,.
\end{equation}
The formulae for the spin-independent cross-section in a general case 
are rather complicated so in order to make some expressions more compact
it is useful to define the following parameters:
\begin{equation}
\label{Ahi}
\mathcal{A}_{h_i}
\equiv
\frac{\tilde{S}_{{h_i}\hat{h}}\left(F^{(N)}_d+F^{(N)}_u\right)
+\tilde{S}_{{h_i}\hat{H}} \left(\tan\beta F^{(N)}_d-\cot\beta F^{(N)}_u \right)
}{\tilde{S}_{h\hat{h}}\left(F^{(N)}_d+F^{(N)}_u\right)
+\tilde{S}_{h\hat{H}} \left(\tan\beta F^{(N)}_d-\cot\beta F^{(N)}_u \right)}
\frac{\tilde{S}_{{h_i}\hat{h}_i}}{\tilde{S}_{h\hat{h}}}
\left(\frac{m_h}{m_{h_i}}\right)^2
\,.
\end{equation}
This is the product of the coupling to a nucleon, the propagator 
and the value of the leading component 
for the scalar $h_i$ divided by the same product for $h$.
Of course, $\mathcal{A}_h=1$ and ${\cal A}_H$ (${\cal A}_s$)
vanishes in the limit $m_H\to \infty$ ($m_s\to \infty$).  
We define also some combinations of the above parameters:
\begin{equation}
\label{Bhi}
\mathcal{B}_{\hat{h}_i}
\equiv
\frac{\tilde{S}_{h\hat{h}_i}}{\tilde{S}_{h\hat{h}}}
+\mathcal{A}_H \frac{\tilde{S}_{H\hat{h}_i}}{\tilde{S}_{H\hat{H}}}
+\mathcal{A}_s \frac{\tilde{S}_{s\hat{h}_i}}{\tilde{S}_{s\hat{s}}}
\,,
\end{equation}
which encode the information on the scalar sector (mixing, masses 
and couplings to the nucleons). 
Using the above definitions we rewrite \eqref{fN} in the form
\begin{align}
f^{(N)}
\approx
\frac{\lambda}{\sqrt{2}}\frac{\alpha_{hNN}}{m_h^2}\tilde{S}_{h\hat{h}}
&\Big\{
\mathcal{B}_{\hat{h}}
N_{15}\left(N_{13}\sin\beta+N_{14}\cos\beta\right)
\nn
&\,\,\,
+\mathcal{B}_{\hat{H}}
N_{15}\left(N_{14}\sin\beta-N_{13}\cos\beta\right)
+\mathcal{B}_{\hat{s}}
\left({N_{13}}{N_{14}}-\frac{\kappa}{\lambda}N_{15}^2\right)
\Big\}.
\label{fN-B}
\end{align}
%

\subsection{Blind spot conditions}
\label{subsec:BS}

The blind spots are defined as those points in the parameter space 
for which the LSP-nucleon cross-section vanishes. From eq.~\eqref{fN-B} 
we obtain the following general blind spot condition
\begin{equation}
\label{BS_general}
\mathcal{B}_{\hat{h}}
N_{15}\left(N_{13}\sin\beta+N_{14}\cos\beta\right)
+\mathcal{B}_{\hat{H}}
N_{15}\left(N_{14}\sin\beta-N_{13}\cos\beta\right)
+\mathcal{B}_{\hat{s}}
\left({N_{13}}{N_{14}}-\frac{\kappa}{\lambda}N_{15}^2\right)
=0
\,.
\end{equation}
This condition simplifies very much for the case of a pure Higgsino 
($N_{15}=0$) or a pure singlino ($N_{13}=N_{14}=0$) LSP. For such pure states 
the blind spot condition reads
\begin{equation}
\label{BS-pure_states}
\mathcal{B}_{\hat{s}}=0
\,.
\end{equation}
For a mixed Higgsino-singlino LSP it is convenient to introduce the parameter
\begin{equation}
\label{eta_def}
\eta
\equiv
\frac{N_{15}(N_{13}\sin\beta+N_{14}\cos\beta)}
{N_{13}N_{14}-\frac{\kappa}{\lambda}N_{15}^2}
\end{equation}
which is totally described by the neutralino sector and the dimensionless 
couplings of the singlet superfield in the superpotential i.e.\ $\lambda$ 
and $\kappa$.\footnote{
Note that in $\mathbb{Z}_3$-NMSSM $\kappa$ controls also the neutralino mass 
parameter.
} 
This parameter vanishes for neutralinos which are pure (Higgsino or singlino) 
states. Its absolute value grows with the increasing admixture of the 
sub-dominant components and has a maximum (or even a pole) for a 
specific highly mixed composition. The position and height of such maximum 
depend on the parameters of the model. Whether there is a pole or a maximum 
depends on the relative signs of some parameters. The details are
given in the Appendix.

The parameter $\eta$ can be used to rewrite eq.~\eqref{BS_general} as
\begin{equation}
\left(\mathcal{B}_{\hat{h}}+\eta^{-1}\mathcal{B}_{\hat{s}}\right)
N_{15}\left(N_{13}\sin\beta+N_{14}\cos\beta\right)
+\mathcal{B}_{\hat{H}}
N_{15}\left(N_{14}\sin\beta-N_{13}\cos\beta\right)
=0
\,.
\end{equation}
After using eqs.~\eqref{Nj3Nj5} and \eqref{Nj4Nj5}, the above general 
blind spot condition may be cast in the form
\begin{equation}
\label{BSeta_mixed-L}
\left(\mathcal{B}_{\hat{h}}+\eta^{-1}\mathcal{B}_{\hat{s}}\right)
\left(\frac{m_{\chi}}{\mu}-\sin2\beta\right)
+\mathcal{B}_{\hat{H}}\cos2\beta
=0
\,.
\end{equation}
For a highly Higgsino-dominated LSP, for which $N_{15}$ and $\eta$ have 
very small values, it is better to rewrite eq.~\eqref{BS_general} as:
\begin{equation}
\label{BSeta_Higgsino}
\left(\eta\mathcal{B}_{\hat{h}}+\mathcal{B}_{\hat{s}}\right)
\left({N_{13}}{N_{14}}-\frac{\kappa}{\lambda}N_{15}^2\right)
+\mathcal{B}_{\hat{H}}
N_{15}\left(N_{14}\sin\beta-N_{13}\cos\beta\right)
=0
\,.
\end{equation}
After applying eqs.~\eqref{Nj3Nj5} and \eqref{Nj4Nj5}, this blind spot 
condition for a highly Higgsino-dominated LSP takes the form
\begin{equation}
\label{BSeta_Higgsino-L}
\left(\eta\mathcal{B}_{\hat{h}}+\mathcal{B}_{\hat{s}}\right)
\left[
\left(1+\left(\frac{m_{\chi}}{\mu}\right)^2\right)
\frac{\sin2\beta}{2}-\frac{m_{\chi}}{\mu}
-\frac{\kappa}{\lambda}
\left(\frac{1-\left(\frac{m_{\chi}}{\mu}\right)^2}{\frac{\lambda v}{\mu}}
\right)^{\!\!2}
\right]
+\mathcal{B}_{\hat{H}}
\frac{1-\left(\frac{m_{\chi}}{\mu}\right)^2}{\frac{\lambda v}{\mu}}\cos2\beta
=0
\,.
\end{equation}
In many cases considered in this paper the contribution from 
$\mathcal{B}_{\hat{H}}$ may be neglected. Then the blind spot conditions 
simplifies to
\begin{equation}
\label{BSeta_BH=0}
\frac{\mathcal{B}_{\hat{s}}}{\mathcal{B}_{\hat{h}}}=-\eta
\,.
\end{equation}

In the rest of the paper we will analyze in some detail the above 
blind spot conditions for several cases and approximations.

\section{Blind spots without interference effects}
\label{sec:fh}

Let us start with a case in which $f_s^{(N)}$ and $f_H^{(N)}$ are negligible
so blind spots correspond to $f_h^{(N)}\approx0$ and result from an 
accidentally vanishing $h\chi\chi$ 
coupling.\footnote{
We do not consider in this paper the possibility that a blind spot may 
originate from vanishing coupling of the Higgs scalar to nucleons, 
i.e.\ vanishing $\alpha_{hNN}$ in eq.~\eqref{fN-B}. In principle, this may
happen if $h$ has a non-zero $\hat{H}$ component and $\tan\beta$ is large 
enough so that the second term in the square bracket of 
eq.~\eqref{alpha-hNN_G} 
for $h_i=h$ cancels the first (usually dominant) term in that bracket.
}
Generically the contributions from $s$ and $H$ exchange 
are very small when these scalars are very heavy. Then, the 
quantities $\mathcal{A}_H$ and $\mathcal{A}_s$ defined in \eqref{Ahi} 
are negligible and eq.~\eqref{Bhi} reduces to 
$\mathcal{B}_{\hat{h}_i}=\tilde{S}_{h\hat{h}_i}/\tilde{S}_{h\hat{h}}$.
The situation is qualitatively different depending on whether the Higgs 
scalar mixes with other scalars or not so we discuss these cases separately 
in the following subsections.

\subsection{Without scalar mixing}
\label{subsec:fh-nomixing}

\begin{figure}[t]
\center
\includegraphics[width=0.48\textwidth]{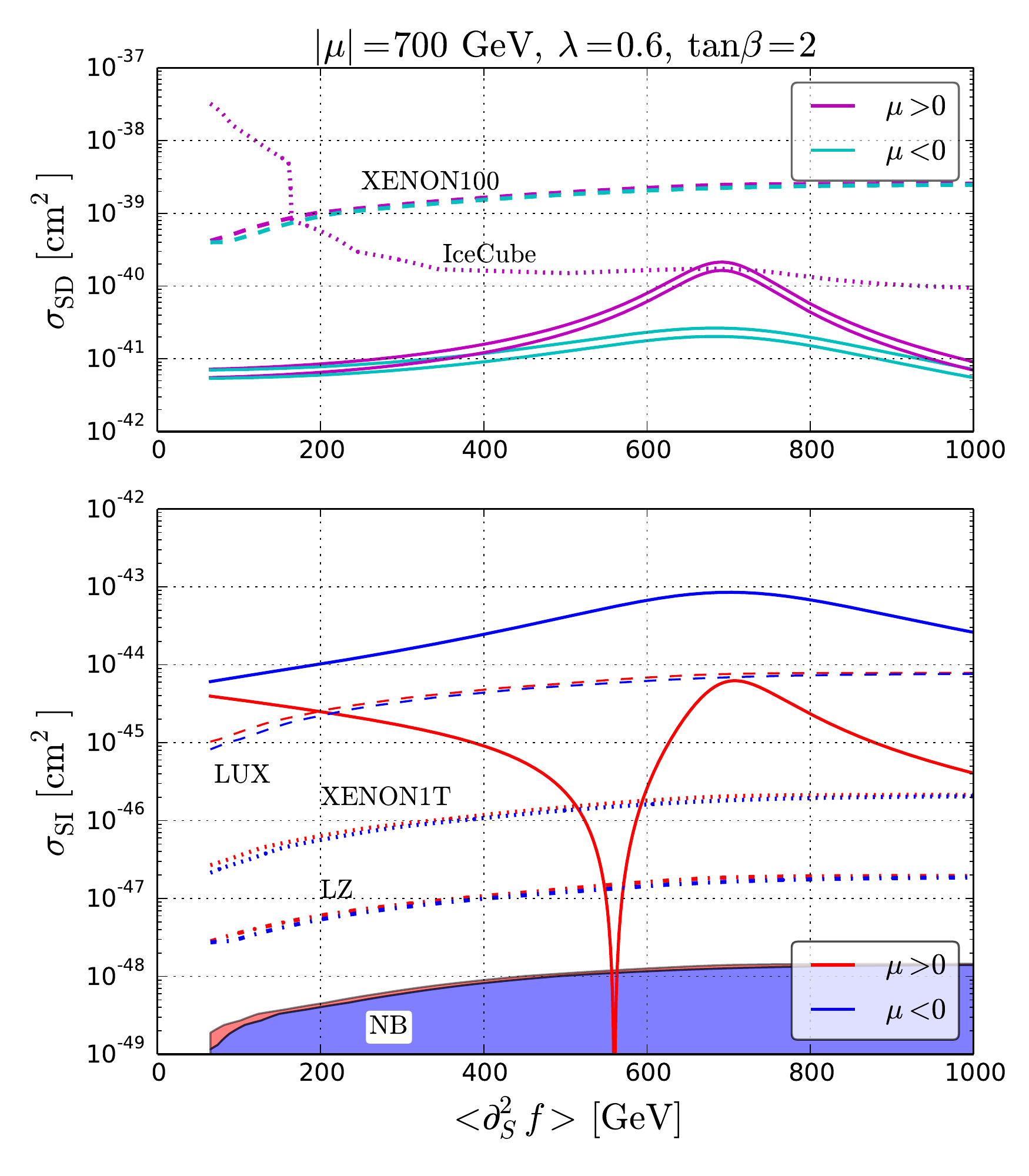}
\includegraphics[width=0.48\textwidth]{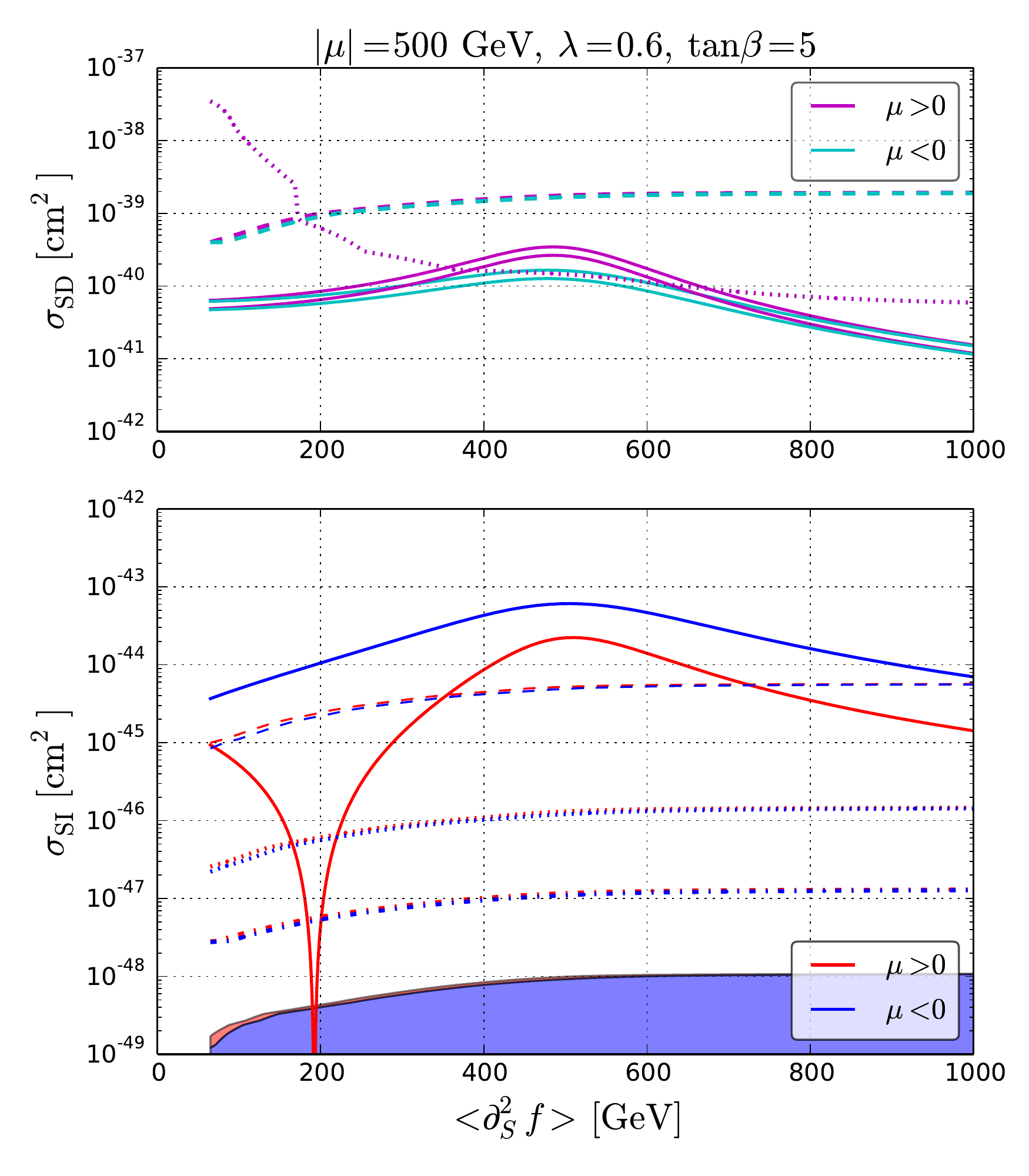}
\caption{
Lower panels: The solid lines show the LSP spin-independent cross-section 
as a function of the diagonal singlino mass term $\langle\partial_S^2f\rangle$ 
for positive (red) and negative (blue) values of parameter $\mu$. 
The dashed, dotted and dashed-dotted lines indicate the corresponding 
upper bounds from, respectively, LUX~\cite{LUX}, XENON1T~\cite{XENON1T} 
and LZ~\cite{LZ} experiments.
The colored areas at the bottom depict the neutrino background (NB)
regions~\cite{NeutrinoB}.
Upper panels: The solid lines show the LSP spin-dependent cross-section 
on neutrons (lower) and protons (upper) 
for positive (purple) and negative (cyan) values of parameter $\mu$. 
The dashed and dotted lines denote the corresponding upper limits 
from, respectively, XENON100~\cite{XENON100} and IceCube~\cite{IceCubeNEW} 
(see details in text).
For all used experimental bounds we assume that the relic density 
of the LSP is equal to the observed value \cite{Planck} 
(otherwise these bounds should be re-scaled by the ratio 
$\Omega_{\rm observed}/\Omega_{\rm LSP}$).
}
\label{fig:blindspot_onlyh}
\end{figure}

Without mixing with (heavy) $\hat{H}$ and $\hat{s}$, the lightest 
scalar $h$ has the same couplings as the SM Higgs. 
In our notation this corresponds to $\mathcal{B}_{\hat{h}}=1$,
$\mathcal{B}_{\hat{H}}=\mathcal{B}_{\hat{s}}=0$. The condition 
\eqref{BS-pure_states} is fulfilled so the SI scattering cross-section 
vanishes when the LSP is a pure singlino or pure Higgsino state.
For a general Higgsino-singlino LSP the amplitude 
\eqref{fN-B} results in the following approximate formula for 
this cross-section:
\begin{equation}
\label{sigma_approx1}
\sigma_{\rm SI}\approx k\cdot 10^{-45}\,{\rm\,cm^2}
\left(\frac{\lambda}{0.1}\right)^2
\frac{N_{15}^2(1-N_{15}^2)}{(0.5)^2}
\end{equation}
where $k$ depends on the value of $\tan\beta$ and typically is of order 
${\mathcal O}(1)$. 
This implies that a highly mixed Higgsino-singlino LSP is strongly constrained 
by the LUX results unless $\lambda$ is very small.
For $\lambda$ which is not small, these constraints may be avoided if there
is some (partial) cancellation between the two terms in the bracket  
multiplying $\mathcal{B}_{\hat{h}}$ in eq.~\eqref{fN-B}
(which results in an unusually small value of $k$ in \eqref{sigma_approx1}).
Such cancellation is equivalent to vanishing of the parameter $\eta$ (defined 
in \eqref{eta_def}) and leads, according to eq.~\eqref{BSeta_BH=0}, 
to a blind spot.
Therefore, highly mixed Higgsino-singlino neutralino dark matter with not very 
small $\lambda$ may be viable only in very special parts of the parameter 
space, close to such blind spots. 
The blind spot condition \eqref{BSeta_mixed-L} for the present 
values of the $\mathcal{B}_{\hat{h}_i}$ parameters, 
$\mathcal{B}_{\hat{h}}=1$, $\mathcal{B}_{\hat{H}}=\mathcal{B}_{\hat{s}}=0$, 
simplifies to:
\begin{equation}
\label{bs_fh_0}
\frac{m_{\chi}}{\mu}-\sin2\beta=0\,.
\end{equation}
This result is analogous to the one obtained in~\cite{Hall} for the 
Higgsino-gaugino LSP in MSSM, but with opposite sign between the two terms 
in the l.h.s.\ This difference stems from the fact that both off-diagonal 
terms, mixing the singlino with two Higgsinos, have the same sign while 
the two analogous terms, mixing any of the gauginos with the Higgsinos, have opposite signs. 
Notice that if $\tan\beta$ is not small, the blind spot condition 
implies a singlino-dominated LSP ($|m_{\chi}|\ll|\mu|$) for which $f_h^{(N)}$ is 
suppressed anyway.
Thus, for a Higgsino-singlino LSP and large $\tan\beta$ this kind of a blind 
spot does not help much in suppression of SI scattering cross-section. 
On the other hand, for small $\tan\beta$ and highly mixed singlino-Higgsino 
LSP the blind spot condition may be satisfied provided that
$\mu\,\langle\partial_S^2f\rangle$ is 
positive\footnote{
As we explained in section \ref{NMSSM}, the sign of $m_\chi$ is the same as 
that of the diagonal singlino entry, $\langle\partial_S^2f\rangle$, in the 
neutralino mass matrix. 
In the scale-invariant NMSSM and in our convention with $\lambda>0$,
the sign of the product $\mu\,\langle\partial_S^2f\rangle$ is the same as 
the sign of $\kappa$.
}. 
This is illustrated in Fig.~\ref{fig:blindspot_onlyh} where the SI scattering 
cross-section is plotted as a function of the diagonal singlino mass term 
$\langle\partial_S^2f\rangle$ (equal to $2\kappa v_s$ in the scale-invariant 
NMSSM) for $\lambda=0.6$, for two values of $\tan\beta$ and for both
signs of $\mu$. 
It can be seen that for small values of $\tan\beta$ (=2 in our example) 
the cross-section is substantially above the LUX 
limit\footnote{
We assume in this work that the relic density of DM is equal to the value 
consistent with the results obtained by the Planck satellite \cite{Planck}. 
If it would not be the case i.e.\ if the relic density would be smaller in 
a specific scenario, the experimental bounds should be appropriately rescaled 
(and hence relaxed).
}
for $\mu\,\langle\partial_S^2f\rangle<0$. As expected, the largest 
cross-section is for $\langle\partial_S^2f\rangle\approx-\mu$ corresponding
to the maximal singlino-Higgsino mixing. Even in the region 
with $\langle\partial_S^2f\rangle$ several times larger than $|\mu|$, 
i.e.\ for a Higgsino-dominated LSP, a small singlino component is  
enough to push the cross-section above the LUX limit. The cross-section 
is below the LUX upper bound only for the LSP with a very tiny Higgsino 
admixture i.e.\ for very large values of $\langle\partial_S^2f\rangle$. 

The situation is drastically different for $\mu\,\langle\partial_S^2f\rangle>0$.
The cross-section is substantially smaller in this case and the LUX limit is 
satisfied for a wide range of values of $\langle\partial_S^2f\rangle$.
One can see that most of this region is within the reach of the XENON1T 
experiment. However, in the vicinity of the blind spot defined by the 
condition \eqref{bs_fh_0} (corresponding to $m_{\chi}=0.8\mu$ for $\tan\beta=2$) 
none of the future SI direct detection experiments will be able to exclude 
(or discover) such a singlino-Higgsino LSP. On the other hand, this region 
may be probed with DM detection experiments sensitive to SD interactions. 
The most stringent model 
independent upper bound on SD cross-section is provided by XENON100 for
neutrons~\cite{XENON100}. The limits on the SD DM-proton cross-section, 
provided by the indirect detection experiment IceCube~\cite{IceCubeNEW}, 
depend strongly 
on assumed dominant annihilation channels of dark matter particles. 
Generically in NMSSM with small $\tan\beta$ and decoupled scalars 
the singlino-dominated 
LSP annihilates mainly into $t\bar{t}$ (if the LSP mass is above the 
top quark mass) while the Higgsino-dominated LSP annihilates mainly 
into $WW$ and $ZZ$ (if kinematically allowed). The IceCube limits for 
DM annihilating dominantly to $WW$, $ZZ$ or $t\bar{t}$ are stronger 
than the XENON100 limits (on SD DM-neutron cross-section) for dark matter 
masses above about 100 GeV~\cite{IceCubeNEW}.  
In the upper panels of Fig.~\ref{fig:blindspot_onlyh} SD cross-sections are
shown with superimposed XENON100 and IceCube limits. 
The IceCube limits are computed assuming the LSP annihilation channels 
as obtained from \texttt{MicrOMEGAs}~\cite{Belanger:2013oya} with the 
spectrum computed by \texttt{NMSSMTools 4.8.2}~\cite{NTools1,NTools2} 
for the model parameters as in Fig.~\ref{fig:blindspot_onlyh} and 
$\kappa=A_\kappa=m_S'^2=\xi_F=0$ as well as $A_\lambda$, $\xi_S$ and $m_3^2$ 
chosen in such~way that $\tilde{S}_{h\hat{s}}\approx 0$, 
$m_{a_1},m_s,m_H\approx 3\,\rm TeV$. 
The SD cross-sections we calculated using 
eqs.~\eqref{eq:sigSD}-\eqref{eq:N13N14diff_lambda} 
(which, as we checked, give results in very good agreement 
with those obtained with the help of \texttt{MicrOMEGAs}). 
Note that for 
$\tan\beta=2$, $\lambda=0.6$ and $|\mu|=700$ GeV in the vicinity of the
SI cross-section blind spot the SD cross-section is not much below the 
current IceCube limit. Since the SD cross-section is larger for larger 
Higgsino-singlino mixing, which is proportional to $(\lambda v / \mu)$, 
the SI blind spot is harder to probe by testing the SD cross-section 
if $\lambda$ is smaller and/or $|\mu|$ is bigger 
(see eq.~\eqref{eq:N13N14diff_lambda}). 
Moreover, for larger $\tan\beta$ the SI blind spot occurs for smaller 
values of $|m_\chi/\mu|$, for which the SD cross-section is smaller 
(because the LSP is more singlino-dominated). Thus, for larger $\tan\beta$ 
smaller values of $|\mu|$ are consistent with the IceCube limits, 
as can be seen from the upper right panel of Fig.~\ref{fig:blindspot_onlyh}. 
We should note also that if LSPs annihilate mainly to $b\bar{b}$, 
which may happen e.g.\ when there is a light sbottom in the spectrum,
the IceCube limits are always weaker than the XENON100 ones. In such a case 
the SI blind spots are much harder to probe via SD detection experiments, 
though not impossible.

We should also comment on the fact that for $\tan\beta=1$ and $m_\chi \mu>0$ 
the blind spot condition \eqref{bs_fh_0} is always satisfied as long as
$|\mu|<\langle\partial_S^2f\rangle$ because in such a case the LSP has a 
vanishing singlino component so $m_\chi=\mu$. Value of $\tan\beta=1$ is 
relevant in the context of $\lambda$SUSY \cite{lambdaSUSY} and will be 
particularly hard to probe because in such situation also SD scattering 
cross-section vanishes, see eqs.~\eqref{eq:sigSD}-\eqref{eq:N13N14diff}.

The properties of the LSP change with the increasing value of $\tan\beta$. 
The difference between values of $\sigma_{\rm SI}$ for two signs of $\mu$ 
decreases. As a result, already for $\tan\beta=5$, a substantial 
part of the parameter space with positive $\mu$ and 
$\langle\partial_S^2f\rangle>|\mu|$ is excluded by the LUX data. 
At the same time, the SI cross-section for negative $\mu$ decreases and 
goes below the LUX upper bound for the LSP with the Higgsino admixture
bigger (i.e.\ for smaller values of $\langle\partial_S^2f\rangle$) 
than in the case of smaller $\tan\beta$. 
What does not change is that there is a blind spot only for positive 
$\mu$. The position of the blind spot moves towards smaller 
$\langle\partial_S^2f\rangle$ corresponding to a more singlino-dominated LSP.

As mentioned before, in our analysis we use the tree-level approximation 
for the SI cross-sections. Inclusion of loop corrections does not affect 
our main conclusion that for $m_\chi \mu>0$ a blind spot for the 
SI cross-section exists. The loop effects may only change slightly the 
position of a given blind spot. The computation of even dominant loop
corrections to the SI cross-section is quite involved. The results 
are known only for neutralinos which are pure interaction eigenstates
\cite{Hisano}.
For a pure Higgsino LSP the radiatively corrected SI cross-section 
is of order $\mathcal{O}(10^{-49})$ cm$^2$ so below the irreducible neutrino
background. One should, however, note that such a small SI cross-section 
is a consequence of quite strong cancellations between contributions 
from several different (gluon and quark, including twist-2) operators, 
some of which contribute as much as $\mathcal{O}(10^{-47})$ cm$^2$.
Computation of the loop corrected SI cross-section for (highly) mixed 
Higgsino-singlino LSP is beyond the scope of this work. 
We conservatively estimate that in such a case the loop correction 
to the tree-level cross-section does not exceed a few times 
$10^{-48}$ cm$^2$ i.e.~the biggest twist-2 operator contribution for a pure 
Higgsino with appropriately reduced couplings to the EW gauge bosons. 
Loop corrections of this size would result in a small shift of the 
position of a blind spot: by less than one per cent in terms of 
$\langle\partial_S^2f\rangle$. 
We checked (using \texttt{MicrOMEGAs}/\texttt{NMSSMTools}) 
that similar size of a shift 
of a blind spot position occurs when the gauginos are
not completely decoupled but have masses of order 2 TeV. 
One should stress that the approximations used in our analysis 
result only in some small uncertainties of the exact positions of 
the blind spots but do not influence their existence.

\subsection{With scalar mixing, {\boldmath $m_s\gg m_h$}}
\label{subsec:onlyfh_mixing}

Next we consider the situation when the contributions to $\sigma_{\rm SI}$
from the exchange of $H$ and $s$ may still be neglected 
($\mathcal{A}_H=\mathcal{A}_s=0$) but the mixing of $h$ with other 
scalars may play some role because now 
$\mathcal{B}_{\hat{h}}=1$, 
$\mathcal{B}_{\hat{H}}=\tilde{S}_{h\hat{H}}/\tilde{S}_{h\hat{h}}$,
$\mathcal{B}_{\hat{s}}=\tilde{S}_{h\hat{s}}/\tilde{S}_{h\hat{h}}$.
The effective LSP-nucleon coupling is obtained by putting 
these expressions for the $\mathcal{B}_{\hat{h}_i}$ parameters into eq.~\eqref{fN-B}.
The fact that $\mathcal{B}_{\hat{H}}$ and $\mathcal{B}_{\hat{s}}$ do not vanish 
implies that in the present case a blind spot may exist for $\eta\neq0$. 
However, as we shall see the blind condition still requires $\eta$ to be 
very small. In the rest of this 
subsection we discuss the blind spot conditions in some interesting limits.

\subsubsection{Purity limits}
\label{subsubsec:fh_mixing_purity}

Before analyzing the general mixed LSP let us discuss limiting cases of a pure 
Higgsino and a pure singlino for which the effective coupling to a nucleon
\eqref{fN-B} simplifies to:
\begin{equation}
f^{(N)}_{h}
\approx
\frac{\alpha_{hNN}}{\sqrt{2}m_h^2}\tilde{S}_{h\hat{h}}\mathcal{B}_{\hat{s}}C
=
\frac{\alpha_{hNN}}{\sqrt{2}m_h^2}C\tilde{S}_{h\hat{s}}
\end{equation} 
where $C$ is equal to $\lambda N_{13} N_{14}$  
($-\kappa N_{15}^2$) for the pure Higgsino (singlino).  
Note that, in contrast to MSSM where the effective tree-level coupling of the
pure Higgsino to a nucleon vanishes \cite{Hall}, the effective coupling 
in NMSSM does not vanish as long as the singlet scalar mixes with the Higgs 
doublet i.e.\ when $\tilde{S}_{h\hat{s}}\ne0$. Similarly, such non-zero 
singlet-Higgs mixing implies a non-vanishing SI scattering cross-section 
also for a pure singlino. Notice that the magnitude of the effective coupling 
of the LSP to nucleons, hence also the SI scattering cross-section, is 
controlled by $\kappa$ for the singlino and by $\lambda$ for the Higgsino.

In order to get a feeling about typical  
(i.e.\ without significant cancellations in the amplitude) 
magnitudes of the SI scattering cross-section it is enlightening to show 
simplified formulae assuming that the $\hat{H}$ component of the 
SM-like Higgs mass eigenstate is 
negligible\footnote{
This approximation is justified since the $\hat{H}$ component modifies 
the Higgs scalar coupling to the bottom quark which is very constrained 
by the LHC data (because modifications of this coupling strongly affect 
the Higgs scalar total decay width and, in consequence, all of the 
Higgs scalar branching ratios). 
}:
\begin{equation}
\label{eq:fh_mix_pureHiggsino}
\sigma_{\rm SI}
\approx
{\cal O}(0.25)\cdot
10^{-45}\,{\rm\,cm^2}\left(\frac{\lambda}{0.1}\right)^2
\left(\frac{\tilde{S}_{h\hat{s}}^2\tilde{S}_{h\hat{h}}^2}{0.1\cdot0.9}\right) 
\end{equation}
for a Higgsino LSP,
\begin{equation}
\label{eq:fh_mix_puresinglino}
\sigma_{\rm SI}
\approx 
{\cal O}(1)\cdot
10^{-45}\,{\rm\,cm^2}\left(\frac{\kappa}{0.1}\right)^2
\left(\frac{\tilde{S}_{h\hat{s}}^2\tilde{S}_{h\hat{h}}^2}{0.1\cdot0.9}\right)
\end{equation}
for a singlino LSP. It is clear from the above formulae that, unless the 
couplings and/or the singlet-Higgs mixing are very small, pure Higgsino and 
singlino neutralino dark matter is generically either excluded by LUX or 
is within the reach of the forthcoming direct detection experiments 
such as XENON1T (so it can be soon found or excluded). 
In particular, for widely considered small $\tan\beta$ and $\lambda\sim 0.6$ 
the SI scattering cross-section for the Higgsino LSP is typically of order 
$10^{-44}\,{\rm cm}^2$, which is above the LUX limit for a wide range of 
its masses.

\subsubsection{General Higgsino-singlino LSP}

For the LSP which is a general Higgsino-singlino mixture there are several 
non-zero contributions to $f_h^{(N)}$ including the one proportional to 
$\mathcal{B}_{\hat{h}}$ (see eq.~\eqref{fN-B}) which on its own leads to 
SI scattering cross-section of order 
$10^{-45}\,{\rm cm}^2$ for $\lambda\approx0.1$, as discussed in 
subsection \ref{subsec:fh-nomixing}. 
Thus, if those contributions add constructively in the amplitude the resulting
cross-section is even bigger. On the other hand, if those contributions add 
destructively a new kind of a blind spot may appear.

In the present case the blind spot condition \eqref{BS_general} can be 
rewritten in the form~\eqref{BSeta_mixed-L} as:
\begin{equation}
\label{bs_fh_mix}
\frac{m_{\chi}}{\mu}-\sin2\beta
=
-\frac{\tilde{S}_{h\hat{H}}}{\tilde{S}_{h\hat{h}}}\cos2\beta
-\frac{\tilde{S}_{h\hat{s}}}{\tilde{S}_{h\hat{h}}}
\eta^{-1}\left(\frac{m_{\chi}}{\mu}-\sin2\beta\right)
\,,
\end{equation}
with $\eta$ given by eq.~\eqref{eta-N15}. Notice that the term in the 
bracket cancels with the same term present in the numerator 
of \eqref{eta-N15}. 
The r.h.s.\ of the above expression quantifies the correction to 
eq.~\eqref{bs_fh_0}, coming from the mixing among scalars. It is tempting 
to check whether adding this correction can change the conclusion of 
subsection \ref{subsec:fh-nomixing}. The first term, proportional to 
$\tilde{S}_{h\hat{H}}$, is typically very small since $\tilde{S}_{h\hat{H}}$ 
is strongly constrained by the LHC measurements of the $hb\ov{b}$ coupling. 
This corresponds to $\mathcal{B}_{\hat{H}}\approx0$. Thus, it cannot change 
qualitatively the conclusions of the case without scalar mixing.  
The situation differs greatly in the case of the second term on the r.h.s.\ 
of eq.~\eqref{bs_fh_mix} which may give important corrections to the simple 
blind spot condition \eqref{bs_fh_0}.

For the discussion of the corrections to the blind spot condition
it is useful to express $\tilde{S}_{h\hat{s}}$ in terms of the NMSSM 
parameters (for $m_s\gg m_h$ assumed in this section):
\begin{equation}
\label{eq:Ssh}
\frac{\tilde{S}_{h\hat{s}}}{\tilde{S}_{h\hat{h}}}
\approx
\lambda v\frac{(\Lambda\sin2\beta-2\mu)}{m_s^2}
\approx
{\rm sgn}(\Lambda\sin2\beta-2\mu)\frac{\sqrt{2|\Delta_{\rm mix}|m_h}}{m_s}
\,.
\end{equation}
In the last approximate equality we introduced  $\Delta_{\rm mix}$, defined as 
\begin{equation}
\label{Deltamix}
 \Delta_{\rm mix} \equiv m_h - \hat{M}_{hh}  \,,
\end{equation}
which parameterizes the correction to the Higgs scalar mass due to its mixing 
with the remaining scalars, mainly with the singlet $\hat{s}$. 
For $m_s> m_h$ this correction is always
negative so its magnitude is desired to be small. Notice that smallness of 
$|\Delta_{\rm mix}|$ usually requires some cancellation between the two terms 
in the bracket (especially for large $\lambda$) in the middle part of  
formula \eqref{eq:Ssh} which implies $\mu\Lambda>0$. Notice also that 
the requirement of small $|\Delta_{\rm mix}|$, say smaller than 
$\mathcal{O}(1)$ GeV, implies 
$({\tilde{S}_{h\hat{s}}}/{\tilde{S}_{h\hat{h}}})\lesssim0.1(m_h/m_s)$. 
Therefore, in order to have a strong modification of the blind spot 
condition, at least one of the other factors in the second term of the 
r.h.s.\ of eq.~\eqref{bs_fh_mix} must 
be much larger than one. This sets the condition for the NMSSM parameter 
space which depends on the composition of the LSP.

Because in the rest of this subsection we will neglect the term proportional 
to $\tilde{S}_{h\hat{H}}$ in \eqref{bs_fh_mix} our blind spot conditions will 
be of the form \eqref{BSeta_BH=0}:
\begin{equation}
\label{bs_fh_mix_eta}
\frac{\tilde{S}_{h\hat{s}}}{\tilde{S}_{h\hat{h}}}
\approx-\eta\,.
\end{equation}
One can see that for small $\hat{h}-\hat{s}$ mixing we demand also small $|\eta|$.  
The dependence of $\eta$ on the LSP composition and mass is explicit in 
eq.~\eqref{eta-N15}. Parameter $\eta$ may be small either because the 
numerator in \eqref{eta-N15} is small or because the denominator is large. 
The first possibility corresponds to the standard blind spot \eqref{bs_fh_0}.
The second possibility requires (at least) one of the terms in the denominator
to be large. 
In the case of a highly mixed LSP $(1-N_{15}^2)/N_{15}=\mathcal{O}(1)$ and the 
denominator may be large only when $|\kappa|\gg|\lambda|$. 
This, however, is limited by the perturbativity conditions. 
Moreover, both sides of eq.~\eqref{bs_fh_mix_eta} must have the same sign
which, using \eqref{eq:Ssh} and \eqref{eta-N15}, gives the condition
\begin{equation}
\sgn\left(\kappa\left(m_\chi-\mu\sin2\beta\right)\right)
=-\sgn(\eta)=\sgn(\tilde{S}_{h\hat{s}})
=\sgn(\Lambda\sin2\beta-2\mu))\,.
\end{equation}
It follows that for $m_\chi\mu<0$ a blind spot is possible only when 
the combination of the parameters 
$\kappa\left(\frac{\Lambda}{\mu}\sin2\beta-2\right)$ is also negative. 
In addition, $|\eta|$ is smaller (i.e.\ better for a blind spot 
with small $|\Delta_{\rm mix}|$) when both terms in the denominator of 
eq.~\eqref{eta-N15} are of the same sign which is the case when 
\begin{equation}
\sgn(\kappa)
=-\sgn\left(\left(1+\left(\frac{m_\chi}{\mu}\right)^2\right)\sin2\beta
-2\frac{m_\chi}{\mu}\right)
\,.
\end{equation}

In the present case with a small value of $\tilde{S}_{h\hat{s}}$ it is easier 
to have a blind spot when the LSP is strongly dominated by the singlino 
(or Higgsino) component because then either $N^2_{15}/(1-N^2_{15})$ or 
$(1-N^2_{15})/N^2_{15}$ in the numerator of \eqref{eta-N15} is large.
Let us now discuss these two situations.

\vspace{1ex}
\noindent
{\bf Singlino-dominated LSP}
\vspace{1ex}

It has been already noted that for pure singlino $\eta$ is exactly zero. 
However, a pure singlino can be obtained only for infinite value of $|\mu|$. 
Very large $|\mu|$ is undesirable for multiple reasons, including naturalness 
arguments. For natural values of $|\mu|$ even if the LSP is singlino-dominated 
some Higgsino component is always present which may have non-negligible 
contribution to $\eta$, hence also to a blind spot condition. Notice
also that for a given value of $\mu$  a minimal value of the Higgsino 
component of the LSP grows with $\lambda$ since the latter controls the 
magnitude of the singlino-Higgsino mixing. In what follows we study the 
impact of a non-zero Higgsino component for the existence of a blind spot.

The blind spot condition \eqref{BSeta_BH=0} with $\eta$ 
given by eq.~\eqref{eta-big_kappa} takes the following form
\begin{equation}
\label{bs_fh_mix_S_singlino}
\frac{m_{\chi}}{\mu}-\sin2\beta
\approx
{\rm sgn}\left(\frac{\Lambda}{\mu}\sin2\beta-2\right)
\frac{\kappa}{\lambda}
\sqrt{
\frac{|\Delta_{\rm mix}|m_h}{m_s^2}\,
\frac{2N_{15}^2}{1-N_{15}^2}\,
\left(1+\left(\frac{m_{\chi}}{\mu}\right)^2-2\frac{m_{\chi}}{\mu}\sin2\beta\right)
}\,.
\end{equation}
For a strongly singlino-dominated LSP its mass $|m_{\chi}|$ is much 
smaller than $|\mu|$ so the first term in the l.h.s.\ of 
the above equation is rather small and the blind spot condition 
without the scalar mixing effects (i.e.\ with the r.h.s.\ neglected) 
can be fulfilled only for appropriately large $\tan\beta$ and  
for positive $m_{\chi}\mu$. 
Now we will check whether the scalar mixing effects may lead to  
blind spots with smaller values of $\tan\beta$ and/or negative $m_{\chi}\mu$. 
Such changes are possible only when the r.h.s\ of \eqref{bs_fh_mix_S_singlino} 
is negative because decreasing of $\tan\beta$ and changing the sign 
of $m_{\chi}\mu$ both give negative corrections to the l.h.s\ of the above 
blind spot condition. This gives the condition 
$(\frac{\Lambda}{\mu}\sin2\beta-2)\kappa<0$. In addition, the absolute value of 
the r.h.s\ of \eqref{bs_fh_mix_S_singlino} should not be very small in order 
to give a substantial modification of the blind spot condition. The 
biggest such value is necessary when one wants simultaneously 
to decrease $\tan\beta$ and have negative $m_{\chi}\mu$. 
Let us now discuss such an extreme modification of blind spots.

\begin{figure}[t]
\center
\includegraphics[width=0.48\textwidth]{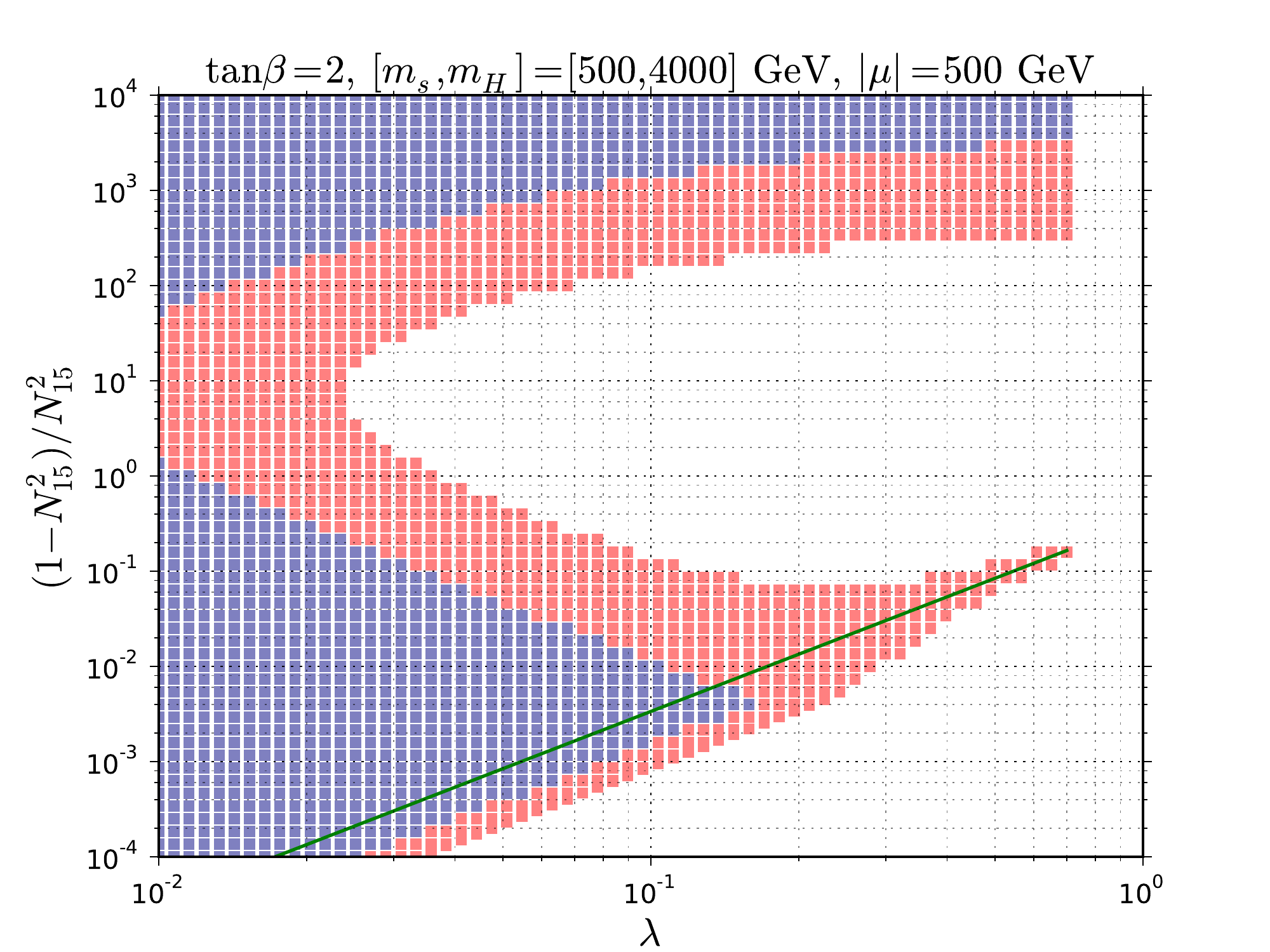}
\includegraphics[width=0.48\textwidth]{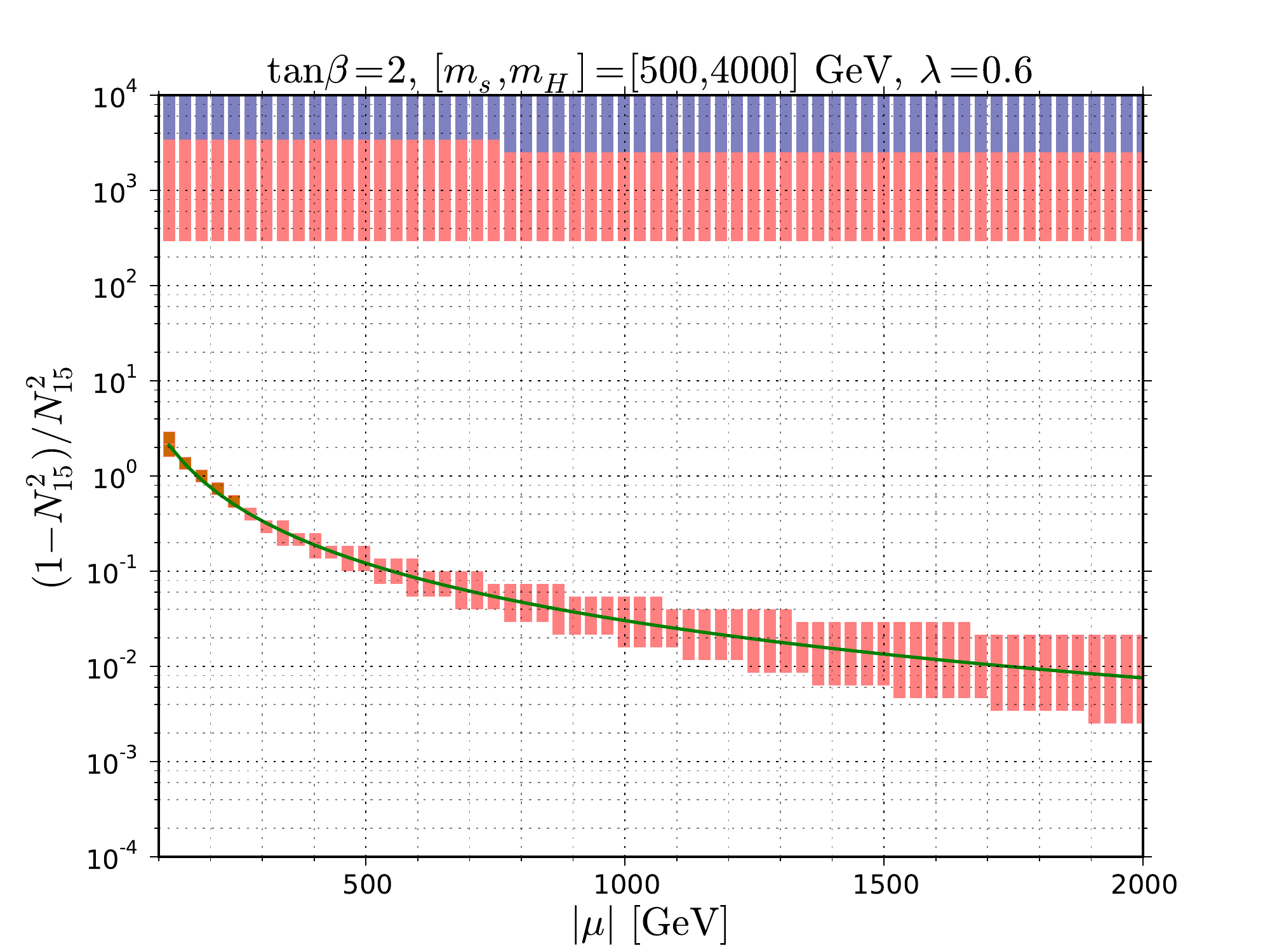}
\caption{Left: Regions of the plane ($\lambda$, $(1-N_{15}^2)/N_{15}^2$) with 
the SI cross-section that can be below the neutrino background for 
$m_{\chi}\mu>0$ (red) and $m_{\chi}\mu<0$ (blue), while keeping 
$10^{-3}\leq|\Delta_{\rm mix}|\leq 1$ GeV and $5\cdot10^{-3}\leq|\kappa|\leq 0.3$. 
Right: The same as in the left panel but as a function of $|\mu|$ and 
fixed $\lambda=0.6$. Green line correspond to the standard blind spot 
condition \eqref{bs_fh_0}. Brown points on the green line for 
$|\mu|\approx120-$250 GeV are excluded by the XENON100 constraints on the 
SD scattering cross-section~\cite{XENON100} (see also 
fig.~\ref{fig:blindspot_hs_scan}). All points are consistent with the 
LHC Higgs data at $2\sigma$.
}
\label{fig:blindspot_onlyh_mixing_scan}
\end{figure}

In the region of large $\lambda\sim0.6$ and small $\tan\beta\sim2$, the 
l.h.s.\ of \eqref{bs_fh_mix_S_singlino} is $\mathcal{O}(1)$ while the r.h.s.\
is generically very small. The reason is that, in addition to the suppression 
by small $|\Delta_{\rm mix}|$, the r.h.s.\ is suppressed also by the factor 
$\kappa/\lambda$ because for $\lambda\sim0.6$ the perturbativity up to the 
GUT scale requires $\kappa\lesssim0.4$ \cite{reviewEllwanger}. The only way 
to enhance the r.h.s.\ would be by the factor $1/\sqrt{1-N_{15}^2}$. However, 
the r.h.s.\ could be of order $\mathcal{O}(1)$ only for extremely pure singlino 
corresponding to $|\mu|\gg\lambda v$. For large $\lambda$ this 
translates to extremely large, hence very unnatural, values of $|\mu|$. 
For example, for $|\kappa|=0.1$, $|\Delta_{\rm mix}|=1$~GeV 
and $m_s=500$ GeV, $|\mu|$ would have to be $\mathcal{O}(20)$ TeV. 
Thus, we conclude that for large $\lambda$ and small $\tan\beta$ it is 
not possible to have a blind spot for a singlino-dominated LSP with 
$m_{\chi}\mu<0$, unless the Higgsino is extremely heavy. For $m_{\chi}\mu>0$ 
such a blind spot can occur only if the standard blind spot condition 
\eqref{bs_fh_0} is approximately satisfied. This can be seen in 
Fig.~\ref{fig:blindspot_onlyh_mixing_scan}  
(in all plots presented in this paper the LEP and LHC Higgs 
constraints (at $2\sigma$ level) are satisfied unless otherwise stated).

The situation changes if $\lambda$ is small. In such a case the r.h.s.\ 
of \eqref{bs_fh_mix_S_singlino} can be enhanced both by $\kappa/\lambda$ 
and by  $1/\sqrt{1-N_{15}^2}$ for not so huge values of $|\mu|$. 
Then a blind spot may appear for $m_{\chi}\mu<0$ and/or small $\tan\beta$ 
provided that at least one of these factors is large enough (of course only 
when $(\frac{\Lambda}{\mu}\sin2\beta-2)\kappa<0$). It can be seen from the 
left panel of Fig.~\ref{fig:blindspot_onlyh_mixing_scan} that for $|\mu|=500$ 
GeV a blind spot with $m_{\chi}\mu<0$ may appear for $\lambda\lesssim0.2$ 
without violating perturbativity constraints. For larger values of $|\mu|$
larger values of $\lambda$ may allow for a blind spot due to decreasing of 
the Higgsino component with increasing $|\mu|$.

We note that it is easier to relax the IceCube constraints on the SD 
cross-section when $|\kappa|$ is not small. This is because for big values of 
$|\kappa|$ the LSP annihilates dominantly (via the s-channel exchange of 
a singlet-like pseudoscalar) into a singlet-like scalar and pseudoscalar 
(if the latter is light enough and LSP has non-negligible singlino component).  
We have verified with \texttt{MicrOMEGAs} that for 
$|\kappa|\sim\mathcal{O}(0.1)$ this is indeed the dominant annihilation 
channel for a singlino-dominated LSP.  The IceCube collaboration 
\cite{IceCubeNEW} does not provide limits on the SD cross-section with 
such an annihilation pattern. It is beyond the scope of the present paper 
to use the IceCube data to accurately calculate limits for such a case. 
However, we expect that such limits would be weaker than for DM 
annihilating into pairs of the SM Higgs bosons because a light singlet-like 
pseudoscalar decays much more often into the bottom quarks and does not decay 
into the gauge bosons. Hence, we expect such limit to be comparable to or 
only slightly better than the one obtained by XENON100.

\begin{figure}[t]
\center
\includegraphics[width=0.48\textwidth]{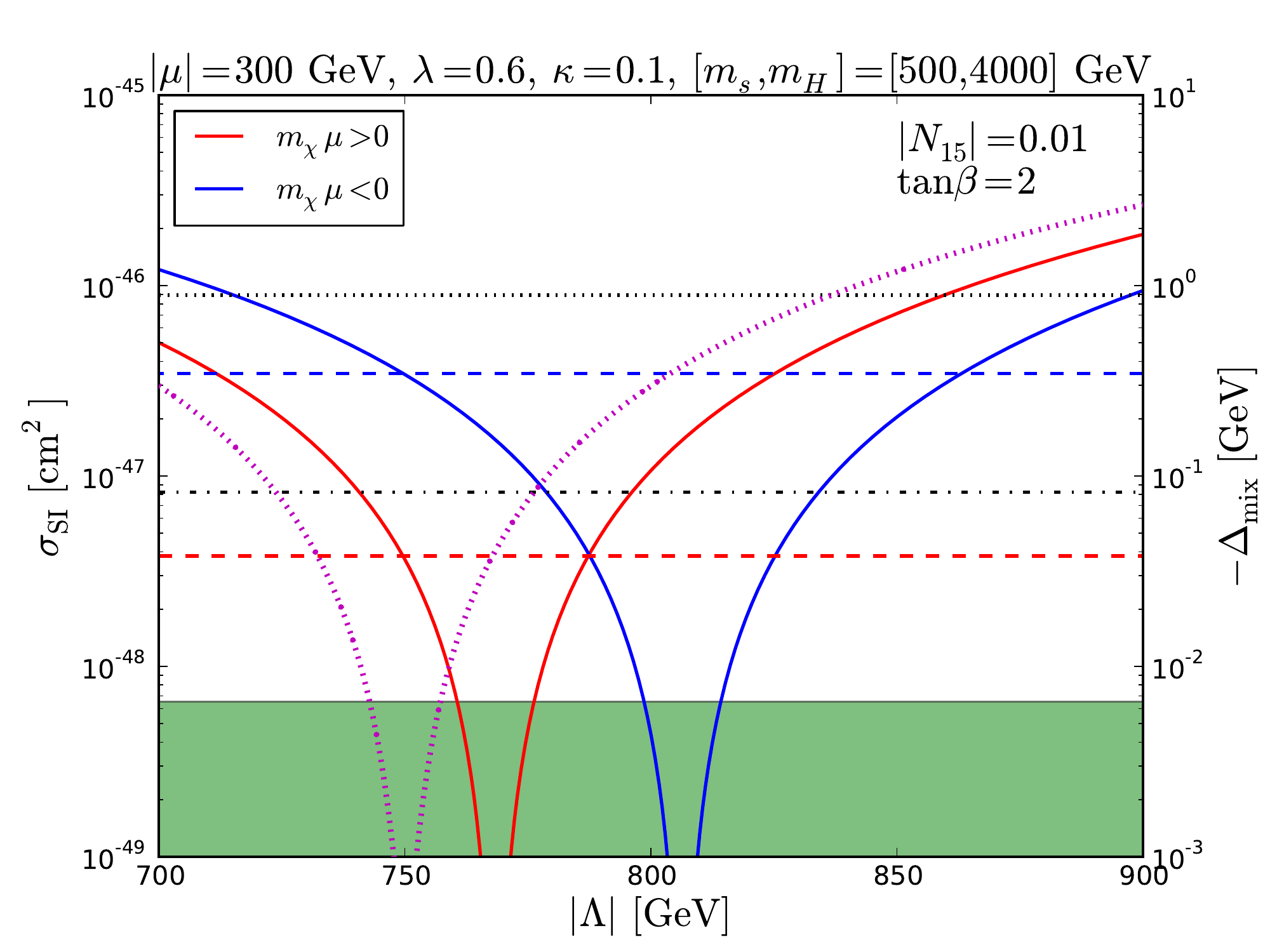}
\includegraphics[width=0.48\textwidth]{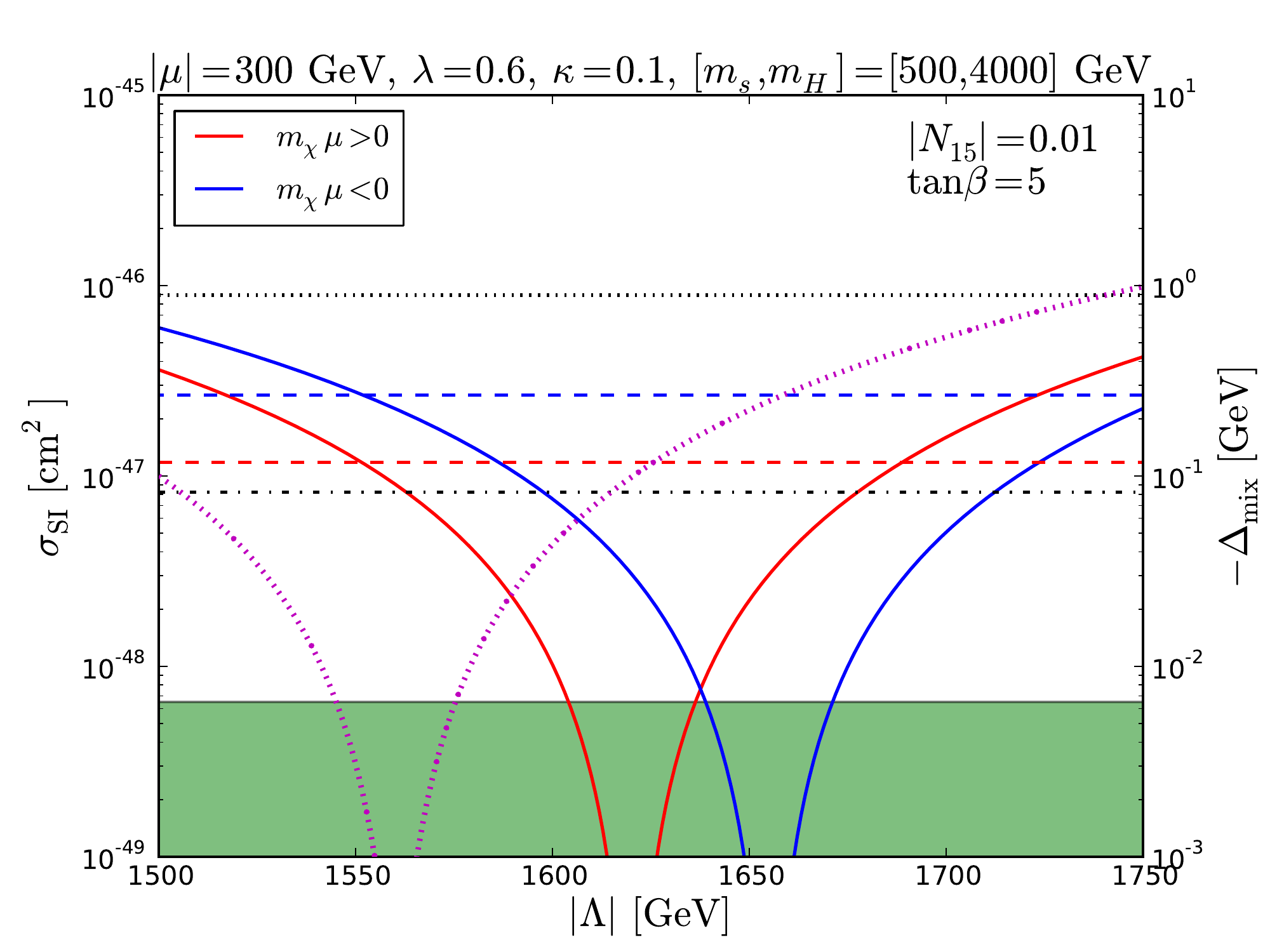}
\caption{Spin-independent scattering cross-section (solid lines) for 
a Higgsino-dominated LSP as a function of $|\Lambda|$ which controls the
size of $|\Delta_{\rm mix}|$ (depicted by coloured dotted curve). 
Parameters $\lambda$, $|\mu|$ and $\tan\beta$ are the same as in
Fig.~\ref{fig:blindspot_onlyh}. Blue and red dashed horizontal lines 
correspond to cross-section with mixing between scalars neglected ($m_s,m_H\to
\infty$), whereas black dotted and dashed-dotted lines denote the XENON1T 
and LZ upper bound, respectively. Green region depicts the neutrino 
background area. The SD cross-section in the vicinity of blind spots is 
below the sensitivity of IceCube 
(independently of the assumed dominant annihilation channel).
}
\label{fig:blindspot_onlyh_mixing_Higgsino}
\end{figure}

\vspace{1ex}
\noindent
{\bf Higgsino-dominated LSP}
\vspace{1ex}

As we discussed in subsection \ref{subsubsec:fh_mixing_purity}, for a pure 
Higgsino the SI cross-section is proportional to the $\hat{h}-\hat{s}$ mixing 
which for $m_s>m_h$ is preferred to be small to avoid large negative 
$\Delta_{\rm mix}$. This implies that for small values of $|\Delta_{\rm mix}|$ 
the LUX constraints on a strongly Higgsino-dominated LSP are generically 
satisfied. However, this is not the case for future direct detection 
experiments so the discussion of blind spots is interesting also in this case.

There are no blind spots for a strongly Higgsino-dominated LSP if the 
contributions from the mixing with $H$ and $s$ scalars are negligible. 
The reason is that for $m_\chi\approx\mu$ the condition \eqref{bs_fh_0} 
could be fulfilled only for $\tan\beta$ very close to 1. 
Let us check whether this conclusion changes after taking into account 
the effects of mixing in the scalar sector.

For a Higgsino-dominated neutralino the second term in the denominator 
in \eqref{eta-N15} may be neglected (unless $\kappa\gg\lambda$). Then, 
substituting \eqref{eq:Ssh} and $\eta$ given by eq.~\eqref{eta-small_kappa} 
into \eqref{BSeta_BH=0}, we get the following blind spot condition
\begin{equation}
\label{bs_fh_mix_S_Higgsino}
\frac{m_{\chi}}{\mu}-\sin2\beta
\approx
-{\rm sgn}\left(\frac{\Lambda}{\mu}\sin2\beta-2\right)
\sqrt{
\frac{|\Delta_{\rm mix}|m_h}{m_s^2}\,
\frac{1-N_{15}^2}{2N_{15}^2}}
\frac{\left(1+\left(\frac{m_\chi}{\mu}\right)^2\right)\sin2\beta-2
\,\frac{m_\chi}{\mu}}
{\sqrt{1+\left(\frac{m_{\chi}}{\mu}\right)^2-2\frac{m_{\chi}}{\mu}\sin2\beta}}
\,.
\end{equation}
For a strongly Higgsino-dominated LSP the ratio $(m_\chi/\mu)^2$ is very 
close to 1 so the numerator of the last factor in the r.h.s\ of the 
above equation is to a very good precision proportional to the combination 
$m_\chi/\mu-\sin2\beta$. So, there are two ways to fulfill the last equation:
either both sides vanish or the factor multiplying $m_\chi/\mu-\sin2\beta$ 
on the r.h.s\ is close to 1. Thus, in the case of a Higgsino-dominated LSP 
there are two kinds of blind spots. First, like in the case without scalar 
mixing, is given by condition \eqref{bs_fh_0} and requires values of 
$\tan\beta$ very close to 1 and $m_\chi$ of the same sign as $\mu$.
The second kind of blind spots is given by the condition
\begin{equation}
\label{eq:bs_fh_mix_S_N15}
1
\approx
\sgn\left(\frac{\Lambda}{\mu}\sin2\beta-2\right)\, 
\frac{\sqrt{|\Delta_{\rm mix}|m_h}}{|N_{15}|m_s}
\frac{1}{\sqrt{1-\sgn(m_\chi\mu)\sin2\beta}}
\,,
\end{equation}
which may be fulfilled only when $(\Lambda/\mu)>(2/\sin2\beta)$.

Notice that for a Higgsino-dominated LSP, i.e.\ small $|N_{15}|$, it follows 
from the last equation that $|\Delta_{\rm mix}|$ is preferred
to be small for a blind spot to occur. Thus, 
the tuning of parameters required to keep $|\Delta_{\rm mix}|$ small
automatically gives some suppression of the SI scattering cross-section, 
provided that $(\Lambda/\mu)>(2/\sin2\beta)$. 
However, the strength of this suppression depends on some other parameters.
For example, for a fixed value of the singlino component in the LSP, $N_{15}$, 
it depends on the sign of $\mu$. This follows from the last factor in the 
r.h.s.\ of eq.~\eqref{eq:bs_fh_mix_S_N15} and 
is illustrated in Fig.~\ref{fig:blindspot_onlyh_mixing_Higgsino} 
for $\lambda=0.6$ and two values of $\tan\beta$. 
The value of $|\Delta_{\rm mix}|$ is bigger when $\mu$ (and in this 
case also $\Lambda$) is negative. As usually, the dependence on the sign 
of $\mu$ is more pronounced for smaller values of $\tan\beta$. 
For $\tan\beta=2$ the value of $|\Delta_{\rm mix}|$ for negative $\mu$ is 
about an order of magnitude bigger than for positive $\mu$. 
So, for a given LSP composition, a blind spot with positive $m_{\chi}\mu$ is 
preferred because it has a bigger Higgs mass. Indeed, it can be seen in 
Fig.~\ref{fig:blindspot_onlyh_mixing_scan} that for $|\Delta_{\rm mix}|<1$ GeV 
and $m_{\chi}\mu>0 $ a larger singlino component of the LSP would be allowed 
if constraints on the SI cross-section would reach the level of the neutrino 
background than for $m_{\chi}\mu<0$. This fact can be understood from 
eq.~\eqref{eq:bs_fh_mix_S_N15}. Moreover, for a given admixture of the 
singlino in the LSP larger values of $\lambda$ would be possible for 
$m_{\chi}\mu>0 $.

Let us also point out that for large $\lambda\sim0.7$, the
perturbativity up to the GUT scale requires $\kappa\lesssim0.3$ which in 
the scale-invariant NMSSM implies that the diagonal singlino mass term 
is smaller than $|\mu|$, hence the LSP would be dominated by the singlino. 
Therefore, the above situation can be realized only in general NMSSM in which 
the LSP can be Higgsino-dominated provided that $\mu'$ parameter
(defined below eq.~\eqref{Lsoft}) is large enough.

\section{\boldmath Blind spots with interference effects between $h$ 
and $H$ exchange}
\label{ref:sec_fhH}

Let us now consider the case in which $f_h^{(N)}$ is not necessarily small but 
interferes destructively with the contribution $f_H^{(N)}$ mediated by the heavy 
Higgs doublet. This kind of blind spots in the context of MSSM was identified 
in \cite{Wagner} and can be realized if $H$ is not too heavy and $\tan\beta$ 
is large. In such a case the coupling of $H$ to down quarks, hence also to 
nucleons, may be enhanced by large $\tan\beta$ which could compensate 
the suppression of $f_H^{(N)}$ by $m_H^{-2}$ resulting in a non-negligible 
$\mathcal{A}_H$ defined in eq.~\eqref{Ahi}. In this section we neglect 
the contribution from the $s$ exchange and set $\mathcal{A}_s$ to zero.

\subsection{Without mixing with singlet}

In the case of negligible mixing of the scalar doublets with the 
scalar singlet the $\mathcal{B}_{\hat{h}_i}$ parameters are given by
\begin{equation}
\mathcal{B}_{\hat{h}}\approx1+\mathcal{A}_H 
\frac{\tilde{S}_{H\hat{h}}}{\tilde{S}_{H\hat{H}}}
\,,\qquad
\mathcal{B}_{\hat{H}}\approx\frac{\tilde{S}_{h\hat{H}}}{\tilde{S}_{h\hat{h}}}
+\mathcal{A}_H 
\,,\qquad
\mathcal{B}_{\hat{s}}\approx0\,.
\end{equation}
The mixing between the doublets is small and may be approximated as
\begin{equation}
\frac{\tilde{S}_{h\hat{H}}}{\tilde{S}_{h\hat{h}}}
\approx
-\frac{\tilde{S}_{H\hat{h}}}{\tilde{S}_{H\hat{H}}}
\approx
\frac{2(M_Z^2-\lambda^2v^2)}{m_H^2\tan\beta}
\,.
\end{equation}
The last equality was obtained under two assumptions: we assumed 
that there is no mixing of the singlet scalar with the 
doublets\footnote{
Quite often the contribution to the $\hat{H}$ component 
of $h$, generated via mixing of both scalar doublets 
with the singlet scalar, is bigger than the contribution 
coming directly from the off-diagonal $M^2_{\hat{h}\hat{H}}$ 
entry \eqref{MhH} in the Higgs mass matrix \eqref{tilde_M^2}.
} 
and that $\tan\beta\gg1$. The former assumption is specific for 
the present subsection. The latter one is necessary because only then 
$f_H^{(N)}$ contribution to $\sigma_{\rm SI}$ can compete with $f_h^{(N)}$ one.
The $\hat{h}$-$\hat{H}$ mixing given by the last equation is suppressed 
by large values of $\tan\beta$ and $m_H^2$. 
This mixing should be small also from the phenomenological point of view. 
A non-negligible $\hat{H}$ component in $h$ results for large $\tan\beta$ 
in strong deviations from the SM predictions of the Higgs scalar branching 
ratios (because of substantial alteration of the Higgs scalar coupling to 
bottom quarks) which is constrained by the LHC Higgs measurements.

When $\mathcal{B}_{\hat{s}}=0$, the blind spot condition \eqref{BSeta_mixed-L} 
can be written as
\begin{equation}
\label{eq:fhH_BS_0}
\frac{m_{\chi}}{\mu}-\sin2\beta
=
-\frac{\frac{\tilde{S}_{h\hat{H}}}{\tilde{S}_{h\hat{h}}}+\mathcal{A}_H}
{1-\mathcal{A}_H\frac{\tilde{S}_{h\hat{H}}}{\tilde{S}_{h\hat{h}}}}\cos2\beta
\,.
\end{equation}
In the case of large $\tan\beta$ and negligible $\hat{h}$-$\hat{H}$ mixing, 
the expression \eqref{Ahi} for $h_i=H$ simplifies to
\begin{equation}  
\label{eq:AHapprox} 
\mathcal{A}_H\approx 
\frac{
 \left(\tan\beta F^{(N)}_d-\cot\beta F^{(N)}_u \right)
}{\left(F^{(N)}_d+F^{(N)}_u\right)
}\left(\frac{m_h}{m_H}\right)^2
\approx
\frac{F_d^{(N)}}{F_d^{(N)}
+F_u^{(N)}}\left(\frac{m_h}{m_H}\right)^2\tan\beta\
\approx \left(\frac{m_h}{m_H}\right)^2\frac{\tan\beta}{2}  
\,.
\end{equation}
Then the blind spot condition \eqref{eq:fhH_BS_0} takes the form
\begin{equation}
\label{eq:BS_AH}
\frac{m_{\chi}}{\mu}-\sin2\beta
\approx \left(\frac{m_h}{m_H}\right)^2\frac{\tan\beta}{2}  
\,.
\end{equation}
This is a similar result to the one obtained in MSSM~\cite{Wagner}, 
but for the singlino-Higgsino LSP, rather than the gaugino-Higgsino one. 
Note, that $\sgn(m_{\chi}\mu)=1$ is required in contrast to MSSM. 
It follows from \eqref{eq:BS_AH} that a non-negligible contribution 
from $\mathcal{A}_H\approx\left({m_h}/{m_H}\right)^2\tan\beta$ leads to 
a bigger Higgsino component of the LSP necessary to obtain a blind spot. 
However, the LHC experiments have set lower mass limits on 
the MSSM-like Higgs bosons, which are stronger for larger $\tan\beta$. 
At large $\tan\beta$, the most stringent constraints on $m_H$ come from 
the ATLAS \cite{ATLAStautau} and CMS \cite{CMStautau} searches in the
$H/A\to \tau\tau$ channel. The results of those searches were interpreted 
in the context of MSSM as constraints on the $m_A$-$\tan\beta$ plane. 
These limits can be applied to NMSSM in generic cases and it is typically 
a good approximation to identify lower limits on $m_H$ for a given $\tan\beta$ 
with the corresponding ones on $m_A$. After taking into account these limits 
one finds generically $\mathcal{A}_H\lesssim\mathcal{O}(0.5)$. In the left 
panel of Fig.~\ref{fig:bs_hH} the black line corresponds to a blind spot 
\eqref{eq:BS_AH} for $\tan\beta=15$ and $m_H=500$ GeV (resulting in 
$\mathcal{A}_H\approx0.5$) which demonstrates that the Higgsino component 
of the LSP at a blind spot with large $\tan\beta$ can be increased when 
effects of the $H$ exchange are not negligible.

We should also comment on the fact that NMSSM provides a framework for relaxing the experimental constraints on $m_H$, hence also on $\mathcal{A}_H$. Namely,
the mass of the MSSM-like pseudoscalar can be very different from $m_H$ if one admits mixing of the MSSM-like pseudoscalar with the
singlet-dominated pseudoscalar (such mixing can be present even if mixing in the CP-even Higgs sector is strongly suppressed). In such a case, the lower mass limit
becomes weaker if the mixing effects push up the MSSM-like pseudoscalar mass substantially above $m_H$. While recasting the LHC constraints on such a scenario
is beyond the scope of this work, it seems viable that this effect may allow for $H$ light enough to have $\mathcal{A}_H\sim\mathcal{O}(1)$. If this is the
case, a blind spot at large $\tan\beta$ would exist also for a highly mixed Higgsino-singlino LSP. 
This would be in contrast to the case with only $h$ exchange for which at large $\tan\beta$ a blind spot cannot exist with $|m_{\chi}|\approx |\mu|$, see
eq.~\eqref{bs_fh_0} and the green line in the left panel of Fig.~\ref{fig:bs_hH}.

\subsection{\boldmath Mixing with singlet, $m_s\gg m_h$}
\label{subsec:fhH_with_mixing}

If the mixing with the singlet scalar is taken into account the parameter 
$\mathcal{B}_{\hat{s}}$ is no longer vanishing. Neglecting the much 
smaller mixing between the two doublets we get
\begin{equation}
\label{B-fhH-mixing}
\mathcal{B}_{\hat{h}}\approx1
\,,\qquad
\mathcal{B}_{\hat{H}}\approx\mathcal{A}_H 
\,,\qquad
\mathcal{B}_{\hat{s}}
\approx
\frac{\tilde{S}_{h\hat{s}}}{\tilde{S}_{h\hat{h}}}
+\mathcal{A}_H \frac{\tilde{S}_{H\hat{s}}}{\tilde{S}_{H\hat{H}}}
\,,
\end{equation}
and
\begin{equation}
\label{AH-fhH-mixing}
\mathcal{A}_H
\approx
\frac{2\tilde{S}_{H\hat{h}}+\tilde{S}_{H\hat{H}}\tan\beta}
{2\tilde{S}_{h\hat{h}}+\tilde{S}_{h\hat{H}}\tan\beta}
\,\frac{\tilde{S}_{H\hat{H}}}{\tilde{S}_{h\hat{h}}}
\left(\frac{m_h}{m_H}\right)^2
\,,
\end{equation}
where we disregarded the difference between $F_d^{(N)}$ and $F_u^{(N)}$
and assumed $\tan\beta\gg1$. 
Adopting these approximations we get the blind spot condition
very similar to that given in eq.~\eqref{bs_fh_mix}, namely:
\begin{equation}
\label{BS_fhH_mix}
\frac{m_{\chi}}{\mu}-\sin2\beta
\simeq
-\mathcal{A}_H\cos2\beta
-\left(\frac{\tilde{S}_{h\hat{s}}}{\tilde{S}_{h\hat{h}}}
+\mathcal{A}_H\frac{\tilde{S}_{H\hat{s}}}{\tilde{S}_{H\hat{H}}}\right)
\eta^{-1}\left(\frac{m_{\chi}}{\mu}-\sin2\beta\right)
\,,
\end{equation}
with $\eta$ given by eq.~\eqref{eta-N15}.
The corrections to~\eqref{bs_fh_mix}, coming from a non-zero 
amplitude $f_H^{(N)}$ and represented by terms proportional to 
$\mathcal{A}_H$, modify both terms in the r.h.s.\ of \eqref{bs_fh_mix}
by shifting the ``small'' components of the Higgs scalar.
Whether one can neglect one of the terms in the first bracket  
in the above equation, 
depends not only on the value of  $\mathcal{A}_H$ but also on the 
sizes of the $\hat{s}$ components in the scalars $h$ and $H$.
The ratio of these two terms can be written as (the dominant components 
$\tilde{S}_{h\hat{h}}$ and $\tilde{S}_{H\hat{H}}$ are not very different from 
1)
\begin{equation}
\label{Shs/AH/SHs}
\frac{\tilde{S}_{h\hat{s}}}{\mathcal{A}_H\tilde{S}_{H\hat{s}}}
{\rm sgn}(2\mu-\Lambda\sin2\beta)
\left(\frac{m_H^2}{m_s^2}-1\right)
\frac{\sqrt{2|\Delta_{\rm mix}|m_h}}{\lambda v}
\frac{m_s}{\Lambda}
\frac{m_H^2}{m_h^2}\,\frac{2}{\tan\beta}
\,.
\end{equation}
In the following, we focus on the case with $m_s\gg m_H$, otherwise 
the assumption of this section that $f_s^{(N)}$ is negligible while 
$f_H^{(N)}$ is taken into account would be typically violated. 
Taking the limit $m_s\to\infty$, while keeping $\Delta_{\rm mix}$ constant 
(by adjusting $\Lambda$ and $\mu$ appropriately), the above ratio blows 
up which means that the contribution proportional to $\mathcal{A}_H$ in 
the second term of the r.h.s. of eq.~\eqref{BS_fhH_mix} 
is negligible. Then, the blind spot condition differs from the one 
\eqref{bs_fh_mix} without $f_H^{(N)}$ only by the term 
$-\mathcal{A}_H\cos2\beta$ which is always positive 
and might be $\mathcal{O}(1)$. In consequence, the contribution from 
$f_H^{(N)}$ makes it harder to obtain a blind spot with $m_{\chi}\mu<0$.

\begin{figure}[t]
\center
\includegraphics[width=0.48\textwidth]{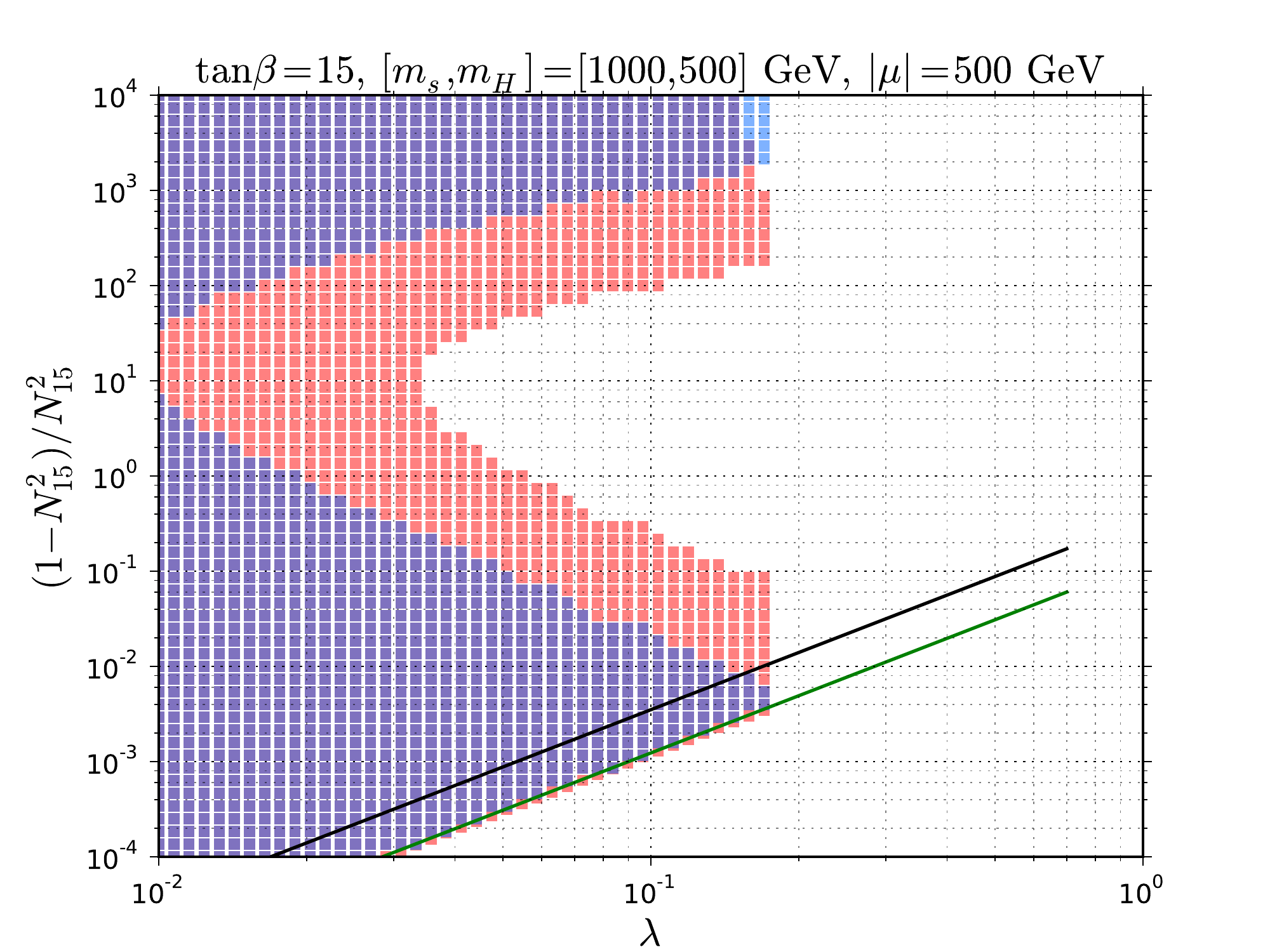}
\includegraphics[width=0.48\textwidth]{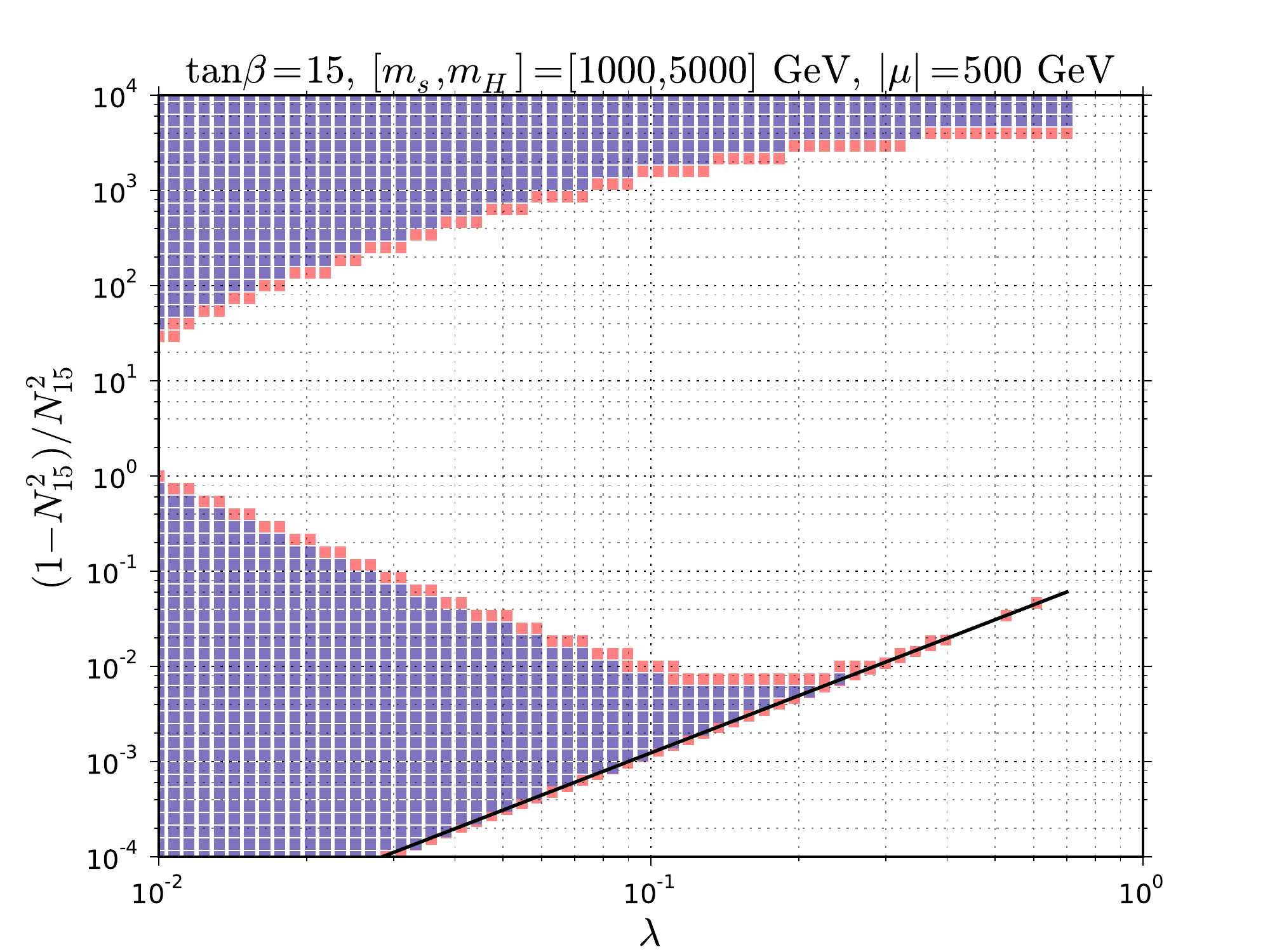}
\caption{The same as in the left panel of 
Fig.~\ref{fig:blindspot_onlyh_mixing_scan} but for 
$\tan\beta=15$ and $m_H=500$ GeV (left) or $m_H=5$ TeV (right). 
Green and black lines correspond to eq.~\eqref{bs_fh_0} and \eqref{eq:BS_AH} 
respectively. All points are consistent with the LHC Higgs data at $2\sigma$.
}
\label{fig:bs_hH}
\end{figure}

Some qualitatively new features may be present only if $m_s$ is in the 
intermediate regime and the ratio \eqref{Shs/AH/SHs} is small. Note that 
the factor in eq.~\eqref{Shs/AH/SHs} involving $|\Delta_{\rm mix}|$ can be 
approximately written as $(0.01/\lambda)\sqrt{|\Delta_{\rm mix}|/(1 {\rm GeV})}$ 
so the ratio \eqref{Shs/AH/SHs} is indeed generically small in the  
phenomenologically most interesting case of small $|\Delta_{\rm mix}|$. 
The ratio could become large only for very small values of $\lambda$ and/or 
for $m_s\gg\Lambda$. Note, however, that under the assumption of small 
$|\Delta_{\rm mix}|$ and large $\tan\beta$ it follows from eq.~\eqref{eq:Ssh} 
that $\Lambda\sim\mu\tan\beta$ so the ratio \eqref{Shs/AH/SHs} is small, 
unless $m_s$ is several orders of magnitude bigger than $|\mu|$. 
This motivates us to assume in the rest of this section the case of 
$m_s\gg m_H$ but with the term proportional to $\tilde{S}_{h\hat{s}}$ 
in the r.h.s.\ of eq.~\eqref{BS_fhH_mix} neglected. 
Then, the blind spot condition can be simplified using: 
\begin{equation}
\label{eq:SHs}
\frac{\tilde{S}_{H\hat{s}}}{\tilde{S}_{H\hat{H}}}
\approx\frac{\lambda v\Lambda}{m_s^2}
\,,
\end{equation}
which is valid as long as $\lambda v\Lambda$ is small in comparison with 
$m_s^2$. From the above equation it should be clear that for large enough 
$\Lambda$ and $\mathcal{A}_H\sim\mathcal{O}(1)$ one can obtain  
$\left|\mathcal{A}_H\frac{\tilde{S}_{H\hat{s}}}{\tilde{S}_{H\hat{H}}}\right|
\gg\left|\frac{\tilde{S}_{h\hat{s}}}{\tilde{S}_{h\hat{h}}}\right|$. 
In such a case the blind spot condition is well approximated by:
\begin{equation}
\label{BS_fhH_mix_AH}
\frac{m_{\chi}}{\mu}-\sin2\beta
\approx
\mathcal{A}_H\left[1-
\frac{\lambda v\Lambda}{m_s^2}
\eta^{-1}\left(\frac{m_{\chi}}{\mu}-\sin2\beta\right)
\right]\,.
\end{equation}
As already noted in the previous subsection, for $m_{\chi}\mu>0$ it is easier 
to have a blind spot for a highly mixed Higgsino-singlino LSP. Indeed, it can 
be seen in Fig.~\ref{fig:bs_hH} that at large $\tan\beta$ with light enough 
$H$ a blind spot is possible for any composition of the LSP. For $m_{\chi}\mu<0$ 
the situation is different. If the mixing in the scalar sector is small, 
only the first term in the square bracket in \eqref{BS_fhH_mix_AH} is relevant 
which makes it harder to obtain a blind spot.  So in order to have  a blind 
spot with $m_{\chi}\mu<0$ the second term in this bracket must be larger 
in magnitude. However, this term may be sizable only for small $|\eta|$, 
i.e.\ for the LSP which is either dominated by singlino or Higgsino. Therefore, 
there are no blind spots for a highly mixed Higgsino-singlino LSP with 
$m_{\chi}\mu<0$. Nevertheless, for large enough $\hat{H}$-$\hat{s}$ mixing 
somewhat bigger Higgsino or singlino component may be possible for large 
$\tan\beta$ if $H$ is light enough, as can be seen from Fig.~\ref{fig:bs_hH}.

Notice, however, that for large $\tan\beta$ and relatively light $H$ the value 
of $\lambda$ exhibits a stronger upper bound. This follows from our requirement 
that negative $\Delta_{\rm mix}$ should have rather small absolute value. Indeed, 
$|\Delta_{\rm mix}|$ is small if $\Lambda\approx\mu\tan\beta$ (in order to 
suppress $M_{\hat{h}\hat{s}}^2$) which results in very large, multi-TeV values 
of $\Lambda$. This in turn implies big $M_{\hat{H}\hat{s}}^2$ unless $\lambda$ 
is small. Nevertheless, the upper bound on $\lambda$ should not be considered 
problematic since there is no strong motivation for big $\lambda$ 
when $\tan\beta$ is large, which is necessary for this kind of a blind spot.

\section{\boldmath Blind spots with interference effects between $h$ 
and $s$ exchange}
\label{ref:sec_fhs}

Now we turn our attention to a case in which the contributions to the 
scattering amplitude from the Higgs scalar and the singlet-dominated scalar are 
comparable. This does not have its analog in MSSM so is particularly 
interesting. In the presence of non-negligible mixing between the singlet 
and the Higgs doublet $f_s^{(N)}$ is generically large if $m_s<m_h$. Light 
singlet-dominated scalar with sizable mixing with the Higgs scalar is particularly 
well motivated since it can enhance the Higgs scalar mass even by 6 GeV 
as compared to the MSSM, allowing for relatively light stops in NMSSM, 
even for large $\tan\beta$ \cite{BaOlPo}.

It was already noticed some time ago \cite{NMSSM_DD} that the contribution 
from the singlet-dominated scalar to the scattering amplitude can be significantly 
larger than the Higgs contribution. Nowadays, such a possibility is excluded 
by the current constraints from the direct detection experiments and it is 
more interesting to study the case in which $f_s^{(N)}$ and $f_h^{(N)}$ are 
similar in magnitude and interfere destructively.\footnote{Such destructive interference was analyzed in some part of the parameter space of the
scale-invariant NMSSM in Ref.~\cite{Cao:2015loa}. }

We neglect the mixing with the heavy scalar $H$ with one exception -- we will 
keep the terms proportional to 
$\left(\tan\beta-\cot\beta\right)\tilde{S}_{{h_i}\hat{H}}$ in~\eqref{alpha-hNN_G} 
for ${h_i}=s,h$.\footnote{
Although, our approach holds for any $\tan\beta$, such terms are crucial 
in the analysis of the possible contribution to the SM-like Higgs scalar 
mass from the mixing with the light singlet-dominated scalar \cite{BaOlPo} 
when $\tan\beta$ is moderate or large.
}
This approximation leads to the following relations
\begin{equation}
\label{gamma}
\frac{\tilde{S}_{s\hat{s}}}{\tilde{S}_{h\hat{h}}} \approx 1
\,,
\qquad\qquad
\gamma\equiv
\frac{\tilde{S}_{h\hat{s}}}{\tilde{S}_{h\hat{h}}}
\approx-\frac{\tilde{S}_{h\hat{s}}}{\tilde{S}_{s\hat{s}}}
\,.
\end{equation}
In the last equation we introduced parameter $\gamma$ which may be 
related to $\Delta_{\rm mix}$ by the following equation
\begin{equation}
\label{Delta_gamma}
\frac{\Delta_{\rm mix}}{m_h}
\approx
1-\sqrt{\frac{1+\gamma^2\left(m_s^2/m_h^2\right)}{1+\gamma^2}}
\approx
\frac12\,\frac{\gamma^2}{1+\gamma^2}\left(1-\frac{m_s^2}{m_h^2}\right)
\,.
\end{equation}
For fixed $m_s$ and small $\gamma$ one gets the proportionality 
$\Delta_{\rm mix}\propto\gamma^2$. 
From \eqref{gamma} we get the following values of the $\mathcal{B}_{\hat{h}_i}$ 
parameters:
\begin{equation}
\mathcal{B}_{\hat{h}}\approx 1-\gamma\mathcal{A}_s\,,\qquad
\mathcal{B}_{\hat{H}}\approx 0\,,\qquad
\mathcal{B}_{\hat{s}}\approx \gamma+\mathcal{A}_s\,.
\end{equation}
Our $\mathcal{A}_s$ parameter can be expressed as
\begin{equation}
\label{eq:As}
\mathcal{A}_s\approx
-\gamma\frac{1+c_s}{1+c_h}\left(\frac{m_h}{m_s}\right)^2
\,,
\end{equation}
where we introduced another convenient parameters
\begin{equation}
\label{c_s}
c_s\equiv1+\frac{\tilde{S}_{s\hat{H}}}{\tilde{S}_{s\hat{h}}}\left(\tan\beta-\frac{1}{\tan\beta}\right)
\,,
\end{equation}
\begin{equation}
\label{c_h}
c_h\equiv1+\frac{\tilde{S}_{h\hat{H}}}{\tilde{S}_{h\hat{h}}}\left(\tan\beta-\frac{1}{\tan\beta}\right)
\,.
\end{equation}
Without mixing with $\hat{H}$ the above quantities would be equal 1. 
In the limit of large $\tan\beta$ the $c_s$ ($c_h$) parameter measures the 
ratio of the couplings, normalized to SM values, of the $s$ ($h$) scalar to 
the $b$ quarks and to the $Z$ bosons. It is easier to make a light scalar 
$s$ compatible with the LEP bounds when $c_s$ is small \cite{BaOlPo}, 
especially for $m_s\lesssim85$ GeV. We should note, however, that $c_s<1$ 
implies $c_h>1$ which in turn leads to suppressed branching ratios of $h$ 
decaying to gauge bosons, so $c_h$ is constrained by the LHC Higgs data.

Note that in contrary to $\mathcal{A}_H$ parameter (see \eqref{AH-fhH-mixing}), 
$\mathcal{A}_s$ can have both signs depending mainly on the sign of $\gamma$. 
LEP and LHC constraints on $\gamma$, ranging from approximately $0.3$ to $0.5$ 
(corresponding to $m_s$ from $m_h/2$ to about 100~GeV), imply that
$|\mathcal{A}_s|\lesssim 1$ (the bound is saturated for $m_s$ around the LEP 
excess).

Because we assumed $\mathcal{B}_{\hat{H}}\approx 0$, the blind spot condition under 
consideration is of the form \eqref{BSeta_BH=0} and reads:
\begin{equation}
\label{eq:fsh_blindspot_G}
\frac{\gamma+\mathcal{A}_s}{1-\gamma\mathcal{A}_s}\approx-\eta
\,.
\end{equation}
It is qualitatively different from the corresponding conditions 
in~\eqref{bs_fh_mix_eta}. The main reason is that the l.h.s.\ of the above 
equation is not generically suppressed (in contrast to the cases considered 
in section \ref{subsec:onlyfh_mixing}). LEP and LHC constraints set upper 
bounds on $|\mathcal{B}_{\hat{s}}/\mathcal{B}_{\hat{h}}|$, nevertheless it can be as large 
as about $0.4$ ($0.3$) for $c_s\approx 1$ ($c_s\approx0$)\footnote{
These upper bounds are quite stable with respect to the change of $m_s$ 
between $m_h/2$ and about 100~GeV.
} 
and therefore could be at least one order of magnitude larger than in the 
case with only $h$ exchange taken into account (see~\eqref{bs_fh_mix_eta}).

\begin{figure}
\center
\includegraphics[width=0.48\textwidth]{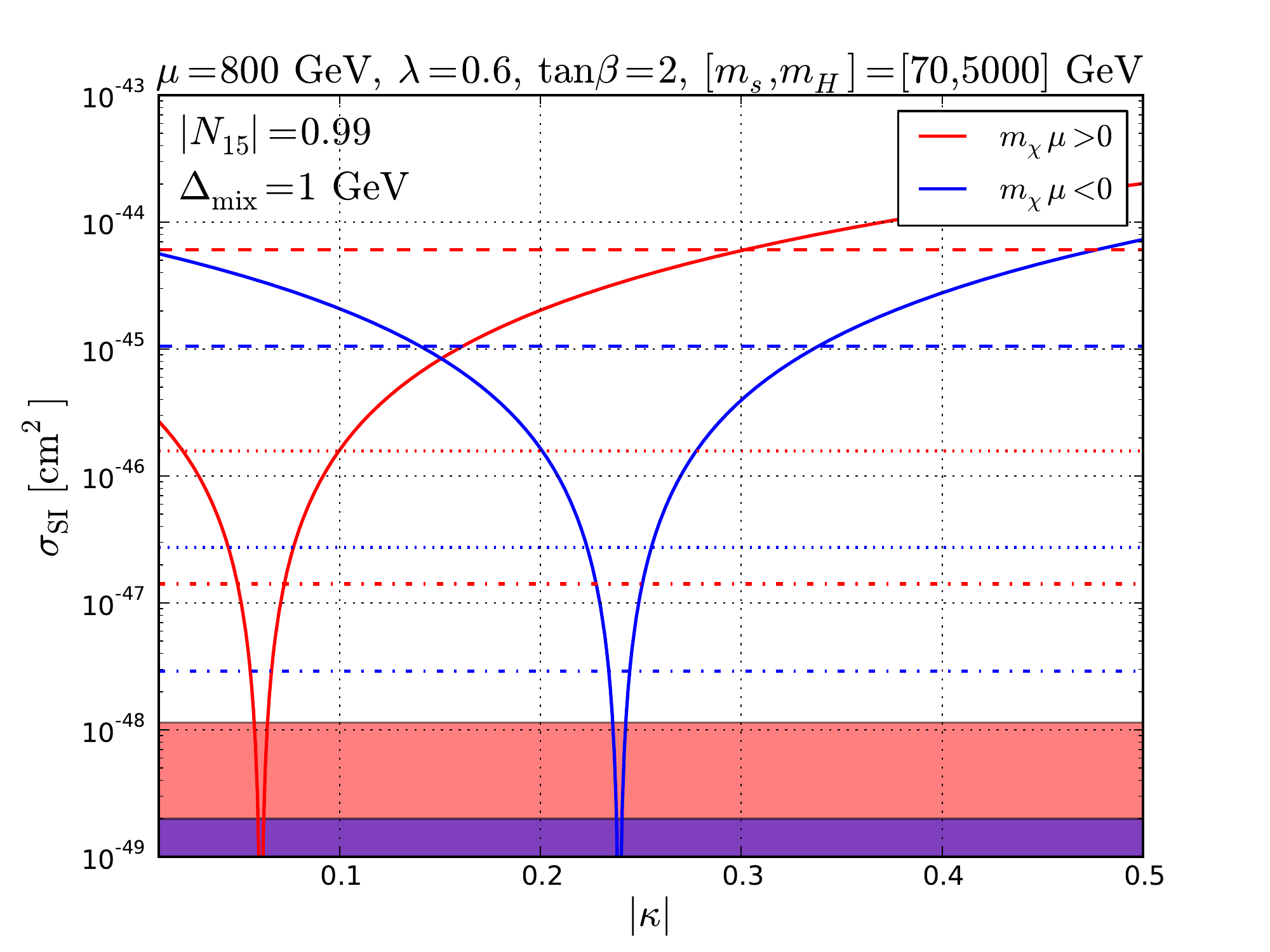}
\includegraphics[width=0.48\textwidth]{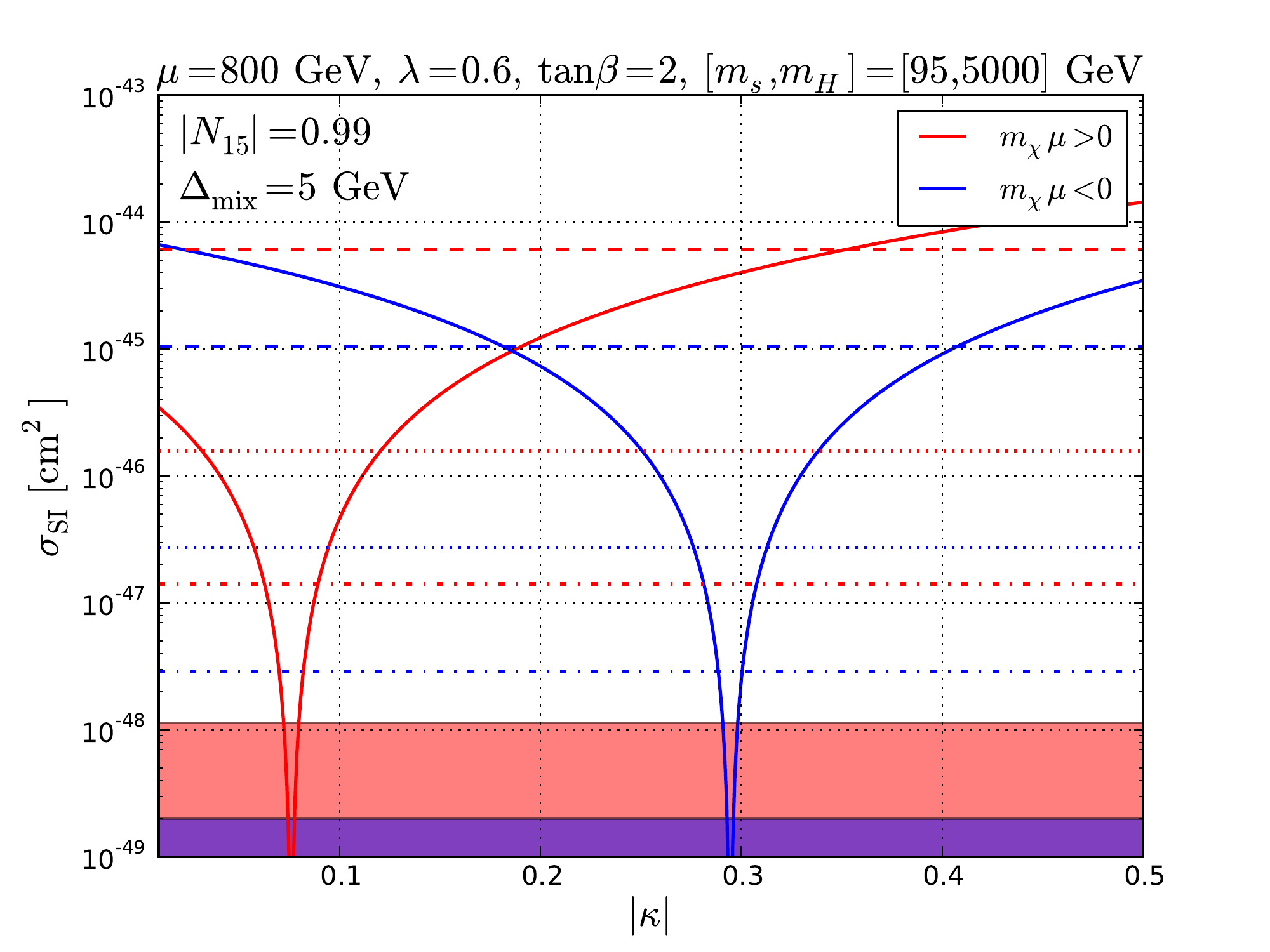}
\caption{
The LSP spin-independent cross-section (solid lines) for $\tan\beta=2$ as 
a function of $\kappa$ which sign is chosen two provide the same signs for 
both sides of~\eqref{bs_fhs_mix}. The horizontal lines show the experimental 
limits as in Fig.\ \ref{fig:blindspot_onlyh}. 
The colored regions depict the corresponding neutrino background levels.
Plots for $\mu<0$ are very similar. The SD cross-section in the vicinity 
of blind spots is below the sensitivity of IceCube 
(independently of the assumed dominant annihilation channel).}
\label{fig:blindspot_hs_singlino_tanb2}
\end{figure}

The above blind spot condition may be rewritten in the form analogous 
to eq.~\eqref{bs_fh_mix}:
\begin{equation}
\label{bs_fhs_mix}
\frac{m_{\chi}}{\mu}-\sin2\beta
\approx
-\frac{\gamma+\mathcal{A}_s}{1-\gamma\mathcal{A}_s}
\eta^{-1}\left(\frac{m_{\chi}}{\mu}-\sin2\beta\right)\,.
\end{equation}
There is one crucial modification as compared to \eqref{bs_fh_mix}: 
\footnote{
The first term in the r.h.s.\ of \eqref{bs_fh_mix} is negligible in any case.
}
\begin{equation}
\label{bs_fhs_mix-modif}
\gamma
\quad\longrightarrow\quad
\frac{\gamma+\mathcal{A}_s}{1-\gamma\mathcal{A}_s}\,.
\end{equation}
Since $|\mathcal{B}_{\hat{s}}/\mathcal{B}_{\hat{h}}|$ does not have to be suppressed it is 
possible to have a blind spot for sizable values of $|\eta|$ independently of 
the sign of $m_\chi\mu$. This implies that a blind spot may occur for larger 
Higgsino-singlino mixing, even for $\lambda$ larger than $|\kappa|$. In
particular, it is now possible to have a blind spot for a singlino-dominated 
LSP for large $\lambda$ and small $\tan\beta$ with sub-TeV $|\mu|$ for both 
signs of $m_\chi\mu$ without violating  perturbativity up to the GUT scale. 
This is demonstrated in Fig.~\ref{fig:blindspot_hs_singlino_tanb2}.
As can be seen for $\lambda=0.6$ and $\tan\beta=2$ the blind spots occur 
for $|\kappa|\lesssim0.4$  (which is necessary to avoid Landau poles below 
the GUT scale for this value of $\lambda$). This is in contrast to the case 
when $\sigma_{\rm SI}$ is dominated by only $h$ exchange, where for a 
singlino-dominated LSP a blind spots with large $\lambda$
and small $\tan\beta$ were present only for $m_\chi\mu>0$.

\begin{figure}[t]
\center
\includegraphics[width=0.48\textwidth]{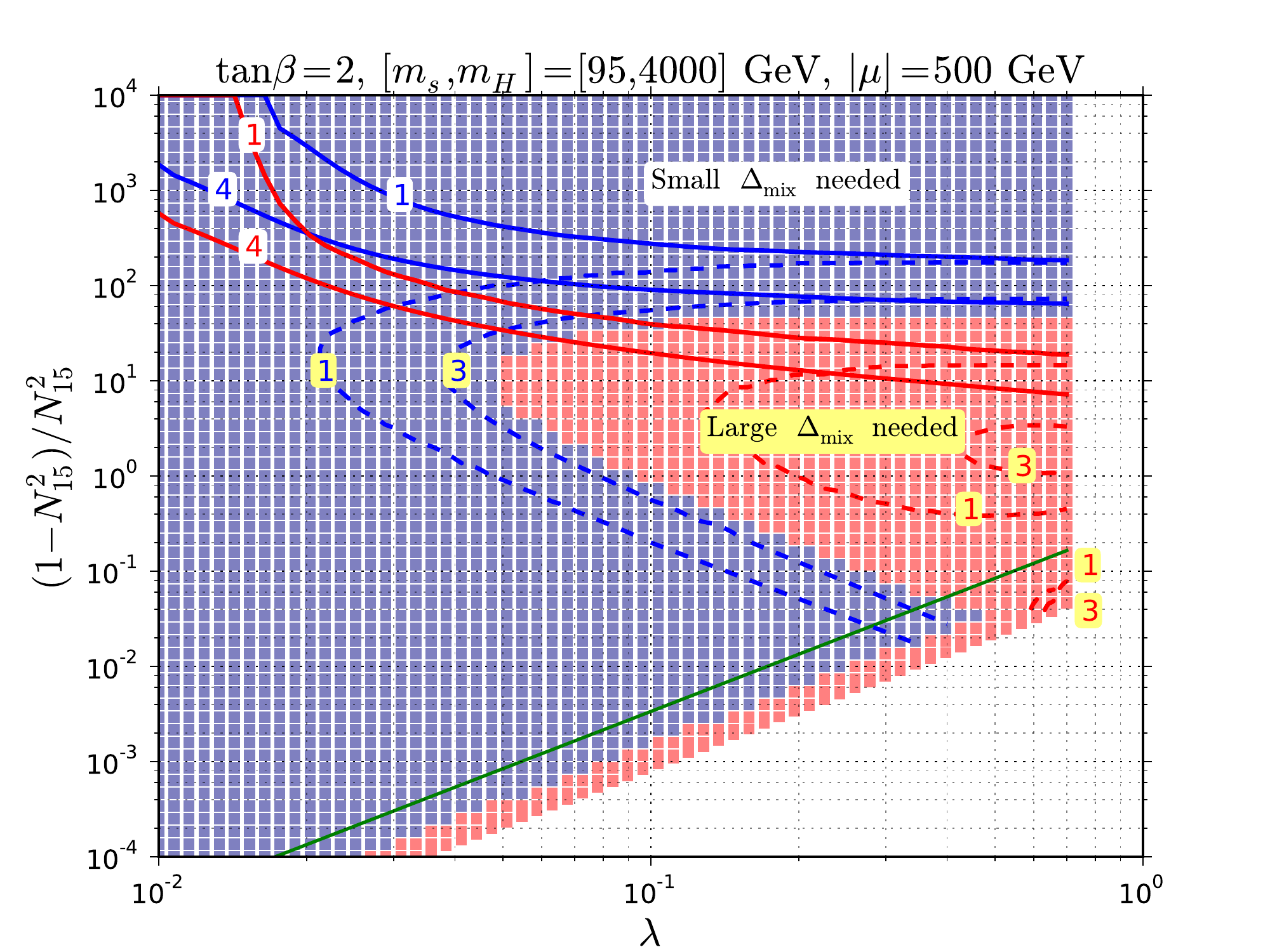}
\includegraphics[width=0.48\textwidth]{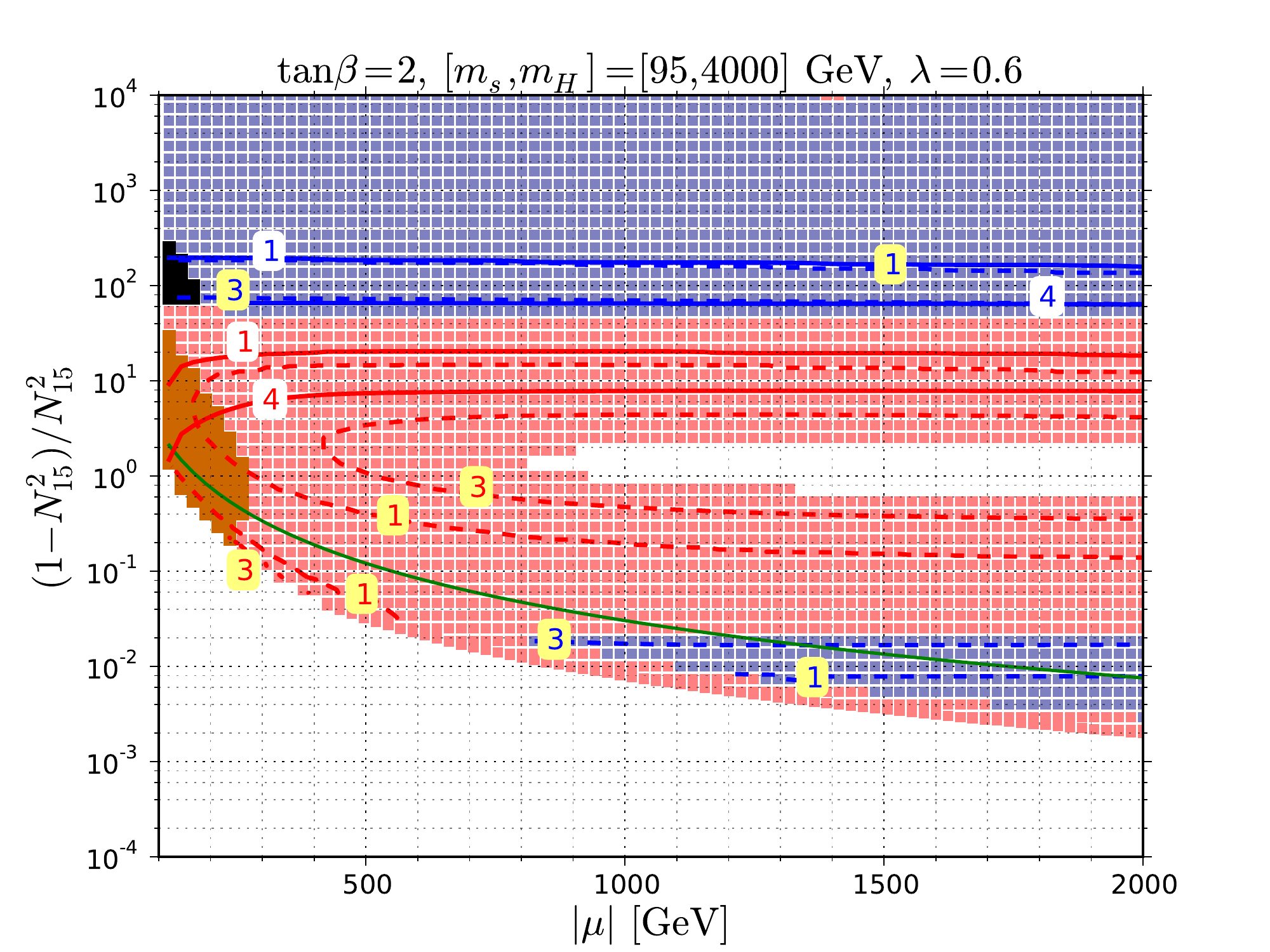}
\caption{Left: Regions of the plane ($\lambda$, $(1-N_{15}^2)/N_{15}^2$) with 
the SI cross-section that can be below the neutrino background for 
$m_{\chi}\mu>0 $ (red) and $m_{\chi}\mu<0$ (blue), while keeping  
$|\kappa|\leq 0.3$ and $|\Delta_{\rm mix}|$ small enough to 
avoid the LEP and LHC constraints and $|\mu|=500$ GeV. The solid contours 
correspond to maximal  value of $\Delta_{\rm mix}$ for which the SI scattering 
cross-section can be below the neutrino background - above these contours 
smaller $\Delta_{\rm mix}$ is required for a blind spot to exist. The dashed 
contours correspond to minimal value of $\Delta_{\rm mix}$ for which the SI 
scattering cross-section can be below the neutrino background - to the right 
of  these contours larger $\Delta_{\rm mix}$ is required for a blind spot 
to exist. Right: The same as in the left panel but as a function of $|\mu|$ 
for $\lambda=0.6$. Black (brown) region is excluded by the XENON100 
constraints on the SD scattering cross-section \cite{XENON100} for 
$m_\chi\mu<0$ ($m_\chi\mu>0$). All points are consistent with the LHC and LEP 
Higgs data at $2\sigma$.
}
\label{fig:blindspot_hs_scan}
\end{figure}

In Fig.~\ref{fig:blindspot_hs_scan} an analogous plots to those presented for 
the heavy singlet case in Fig.~\ref{fig:blindspot_onlyh_mixing_scan} are
shown. It can be seen that, if the singlet-dominated scalar is light, blind spots 
can exist for large $\lambda$ and $\tan\beta=2$ without violating the 
perturbativity bounds for $m_\chi\mu>0$ for (almost) any composition of the 
LSP. The case of $m_\chi\mu<0$ is also less constrained. Nevertheless, if the 
LSP is not Higgsino-dominated, blind spots can exist for large $\lambda$ and 
$m_\chi\mu<0$ only for small range of $N_{15}$ (if $\kappa$ is kept in the 
perturbative regime). For low $\tan\beta$ the most interesting region is for 
large $\lambda$ so in the right panel of Fig.~\ref{fig:blindspot_hs_scan} 
we plot the regions where a blind spot can occur for fixed $\lambda=0.6$ 
as a function of $|\mu|$. It can be seen that for $m_\chi\mu<0$ a blind spot 
can occur for a singlino-dominated LSP if $|\mu|\gtrsim800$ GeV, and the range 
of possible values of $N_{15}$ grows with increasing $|\mu|$. 
For $m_\chi\mu>0$ almost any LSP composition allows for existence of a blind
spot except for some region of a strongly-mixed Higgsino-singlino LSP with 
$|\mu|\gtrsim800$~GeV (in that region a blind spot cannot occur because 
$|\eta|$ is too large to satisfy the blind spot eq.~\eqref{BSeta_BH=0} when 
the precision Higgs data, constraining the Higgs-singlet mixing, are taken 
into account).

The fact that the blind spots can now occur for large $\lambda$ and small 
$\tan\beta$ for much wider range of the LSP composition is not only due to 
the fact that the singlet-dominated scalar is light but also because of 
large Higgs-singlet mixing, hence also large $\Delta_{\rm mix}$. This is 
demonstrated by dashed contours in Fig.~\ref{fig:blindspot_hs_scan} which 
correspond to minimal value of $\Delta_{\rm mix}$
\footnote{
Note that the results in Fig.~\ref{fig:blindspot_hs_scan} come from a scan 
of four parameters: $N_{15}$, $\lambda$, $\Delta_{\rm mix}$ and $\kappa$.
Therefore, for a given point in the $N_{15}$-$\lambda$ plane there might 
be several solutions with the SI scattering cross-section below the 
neutrino background with different values of $\Delta_{\rm mix}$.  
}
for which the SI scattering cross-section may be below the neutrino 
background. It follows from the comparison of these contours with the plot 
in Fig.~\ref{fig:blindspot_onlyh_mixing_scan} (for heavy singlet) that 
$\Delta_{\rm mix}$ above few GeV is required to significantly extend the range 
of the LSP composition for which a blind spot can occur when $\lambda$ is 
large.

It is also interesting to check what happens if one demands large 
$\Delta_{\rm mix}$ so that the Higgs scalar mass gets substantial enhancement 
from the Higgs-singlet mixing effects. In Fig.~\ref{fig:blindspot_hs_scan} 
we also present solid contours that correspond to maximal value of 
$\Delta_{\rm mix}$ for which the SI scattering cross-section may be below the
neutrino background. It can be seen that if one demands $\Delta_{\rm mix}$ 
as small as 1 GeV then for large $\lambda$ there are no blind spots for 
the LSP strongly dominated by the Higgsino component. This can be understood 
in the following way. For large $\Delta_{\rm mix}$ and light singlet 
$|\mathcal{B}_{\hat{s}}/\mathcal{B}_{\hat{h}}|$ is no longer close to zero so 
in order for the blind spot to occur $|\eta|$ should not be close to zero. 
One can see from definition \eqref{eta_def} that 
$|\eta|\sim |N_{15}|$ for the Higgsino-dominated case so a lower bound on 
$\Delta_{\rm mix}$ sets a lower bound on the singlino component of the LSP. 
Noting that $|\mathcal{B}_{\hat{s}}/\mathcal{B}_{\hat{h}}|$ is in a good 
approximation proportional to $\sqrt{|\Delta_{\rm mix}|}$, we conclude that 
a lower bound on $N_{15}^2$ scales proportionally to $\Delta_{\rm mix}$. 
This is in agreement with the results in Fig.~\ref{fig:blindspot_hs_scan}.

Since in this case a SI cross-section blind spot can occur also for 
a highly-mixed Higgsino-singlino LSP one may expect to probe this region 
with SD direct detection experiments. Indeed, XENON100 limits exclude some 
part of the parameter space with SI cross-section blind spots for large 
$\lambda$ and small $|\mu|$ (black and brown points in 
Fig.~\ref{fig:blindspot_hs_scan}). In this region of the parameter space 
the LSP annihilates dominantly to a light singlet-like scalar and 
a pseudoscalar, that typically decay to pairs of bottom quarks so the 
IceCube limits are not expected to be stronger than the XENON100 ones.

We should emphasize that the effect of large Higgs-singlet mixing has 
particularly important implications for models with $\mu'=0$
(i.e.~with vanishing quadratic term in $f(S)$), including the
$Z_3$-invariant NMSSM, because in those models the LSP composition is related 
to the ratio $\kappa/\lambda$. Namely, the LSP is
singlino-dominated if $\lambda>2|\kappa|$. This implies that for large 
$\lambda$, the LSP is typically singlino-dominated and can be highly mixed
Higgsino-singlino only if $|\kappa|$ is close to the upper bound from 
the requirement of perturbativity up to the GUT scale. In consequence, 
in this class of NMSSM models with large $\lambda$ and small $\tan\beta$ 
a blind spot may occur only for 
$2\kappa/\lambda \approx m_{\chi}/\mu\approx\sin(2\beta)$ if the
Higgs-singlet mixing is small. On the other hand, for large Higgs-singlet 
mixing a blind spot can occur for much wider range of $\kappa/\lambda$
(corresponding to different LSP compositions) for $m_{\chi}\mu>0$, while 
for $m_{\chi}\mu<0$ existence of a blind spot may be possible provided that
$|\mu|$ is large enough.

\subsection{\boldmath Large $\tan\beta$ region}

In models with large $\tan\beta$, couplings of $s$ and $h$ scalars to $b$ 
quarks may significantly deviate from the couplings to the massive gauge 
bosons which has important consequences for the SI scattering cross-section. 
From our perspective the most interesting situation takes place when 
$\Delta_{\rm mix}$, being now positive, is large. As stated above, for
$m_s\lesssim 85$~GeV small $|c_s|$ and hence large $\tan\beta$ and small 
$\lambda$ are preferred~\cite{BaOlPo}. For definiteness, let us consider 
$\tan\beta=10$, $\lambda=0.1$ and two representative values of $m_s$, 
70 and 95 GeV, for which the LEP bounds are, respectively, quite severe and 
rather mild. In Fig.~\ref{fig:bs_fhs_N15_delta} we present the points 
(for a few values of $c_s$) for which $\sigma_{\rm SI}$ is smaller than the 
neutrino background for two signs of $m_\chi\mu$. The most apparent difference 
between $c_s>1$ and $c_s<1$ is that in the first case there are no points 
with a Higgsino-dominated LSP, whereas in the second one there is a negative 
correlation between Higgsino admixture and $\Delta_{\rm mix}$ 
(for $N_{15}^2\lesssim0.1$). In order to explain this behavior we rewrite 
the blind spot condition~\eqref{bs_fhs_mix} in the form adequate for the 
Higgsino-dominated limit i.e.\ for $|m_\chi/\mu|\to 1$. The result reads:
\begin{figure}[t]
\center
\includegraphics[width=0.49\textwidth]{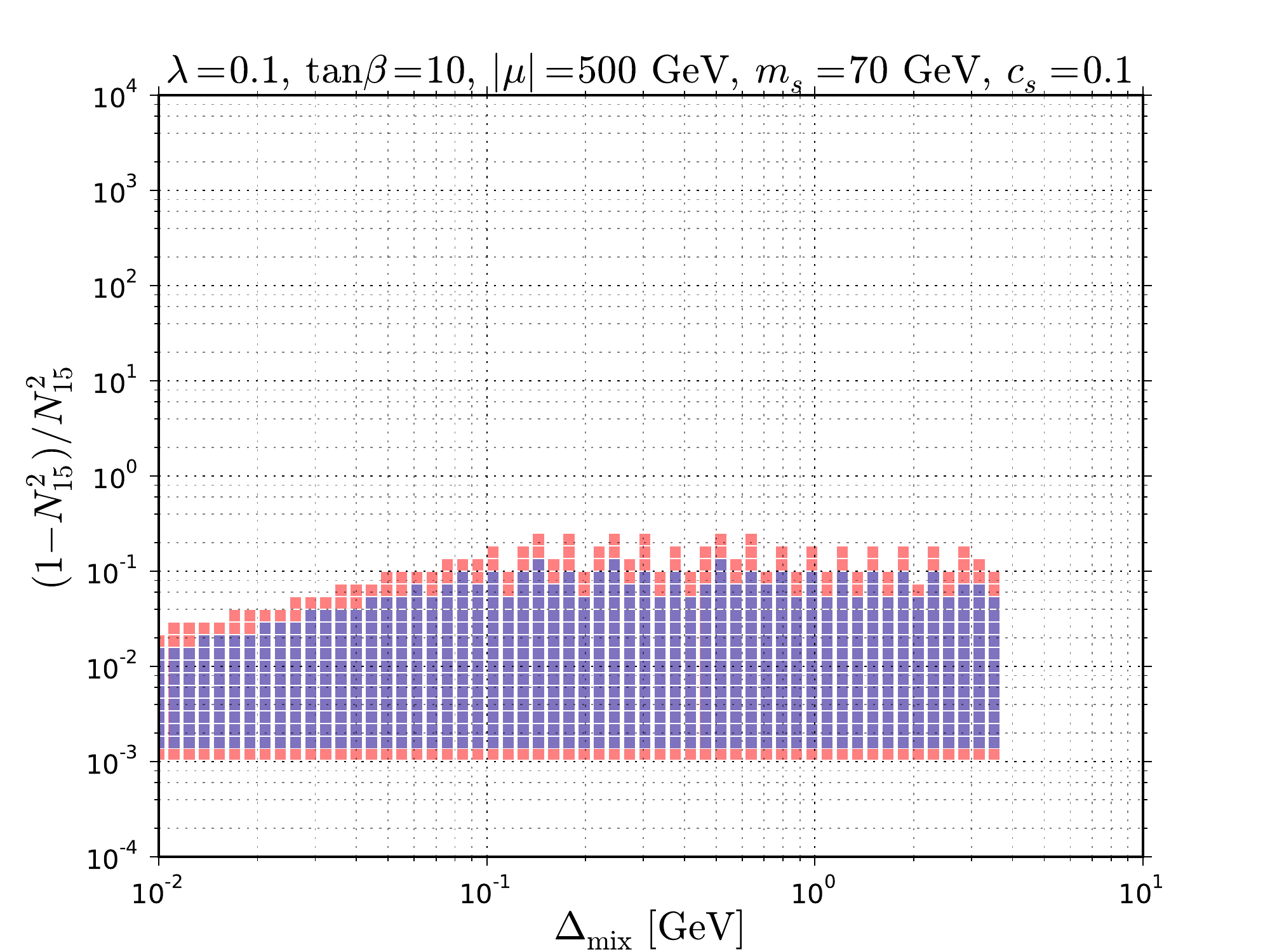}
\includegraphics[width=0.49\textwidth]{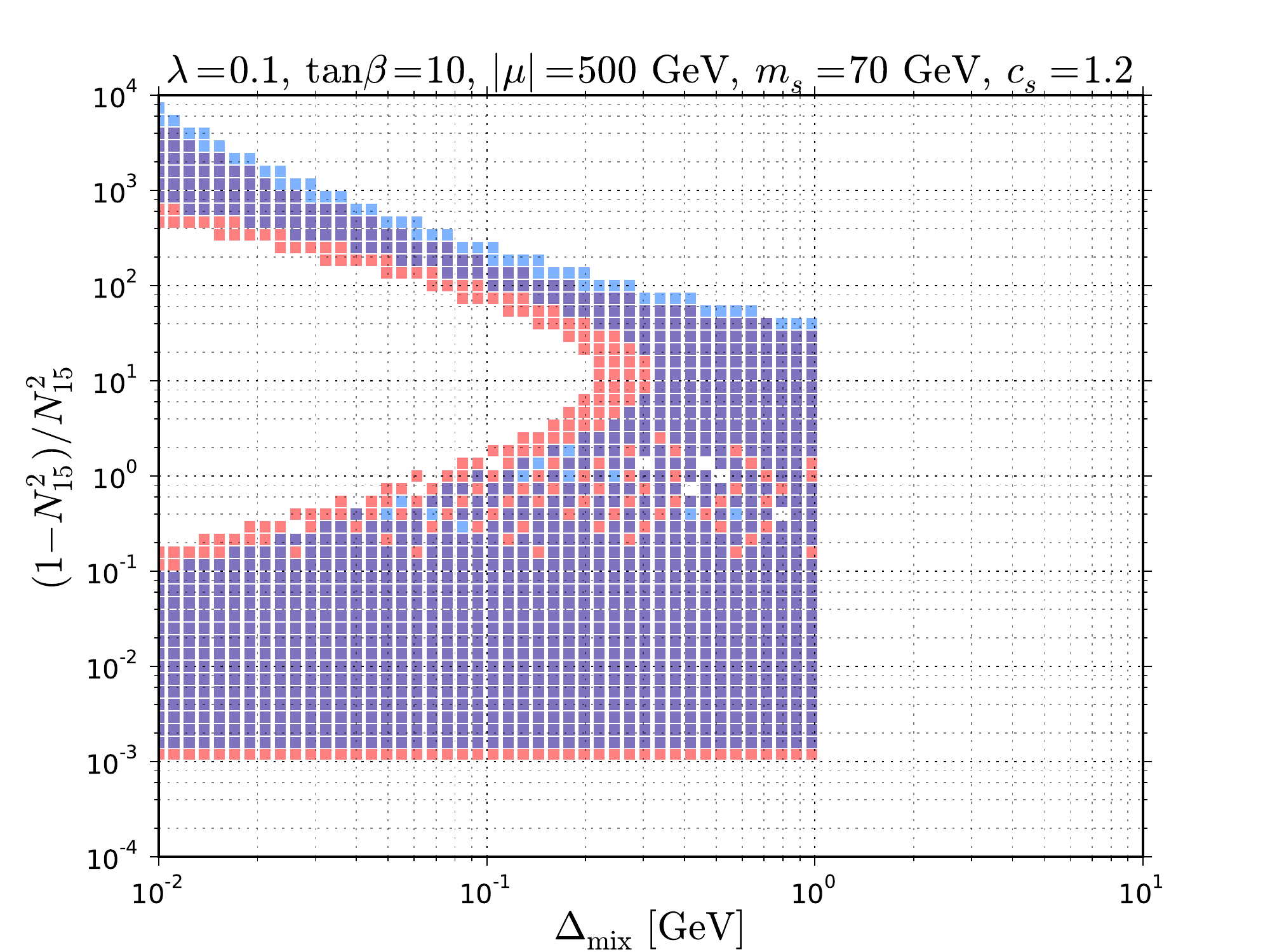}\\
\includegraphics[width=0.49\textwidth]{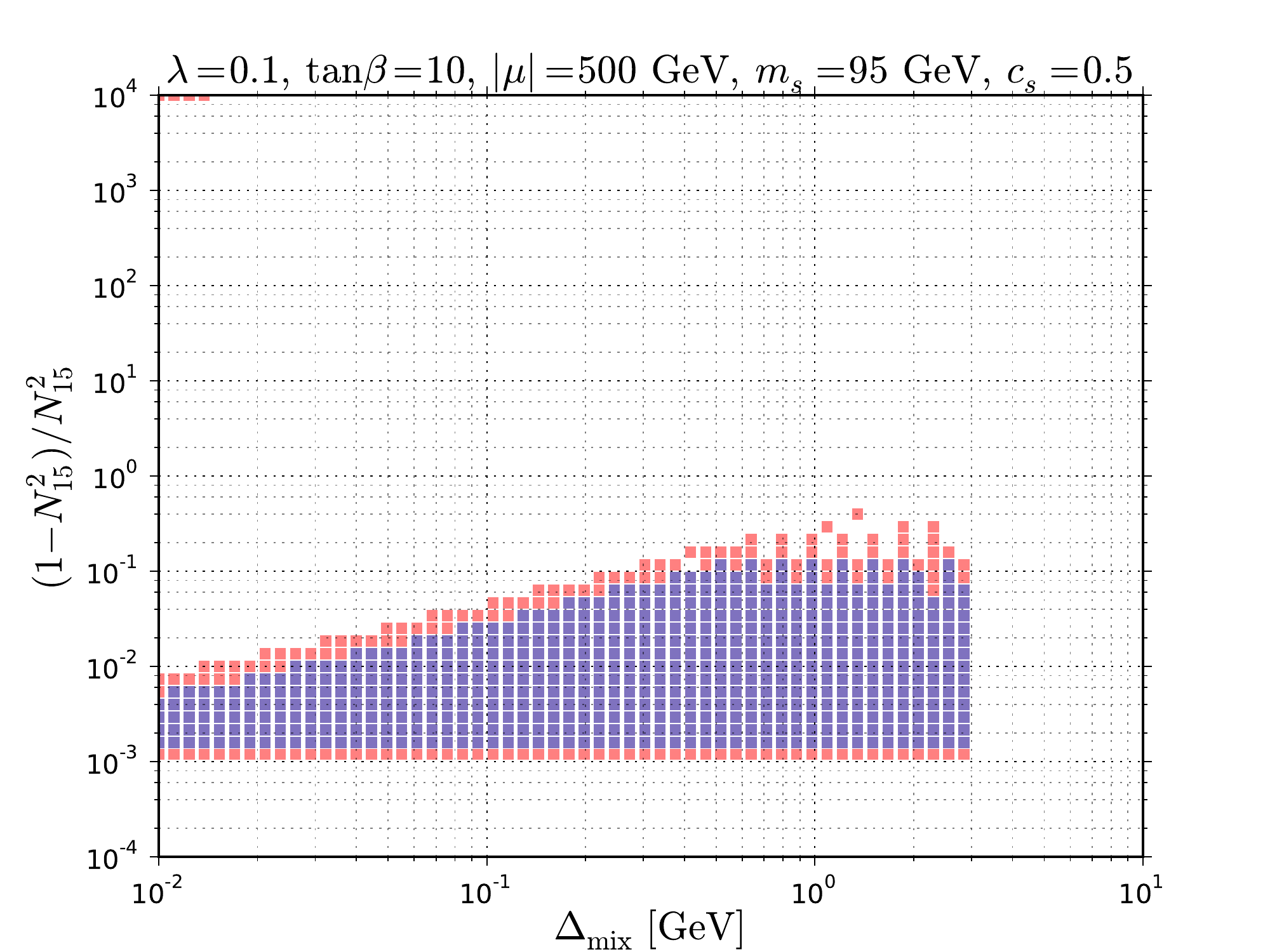}
\includegraphics[width=0.49\textwidth]{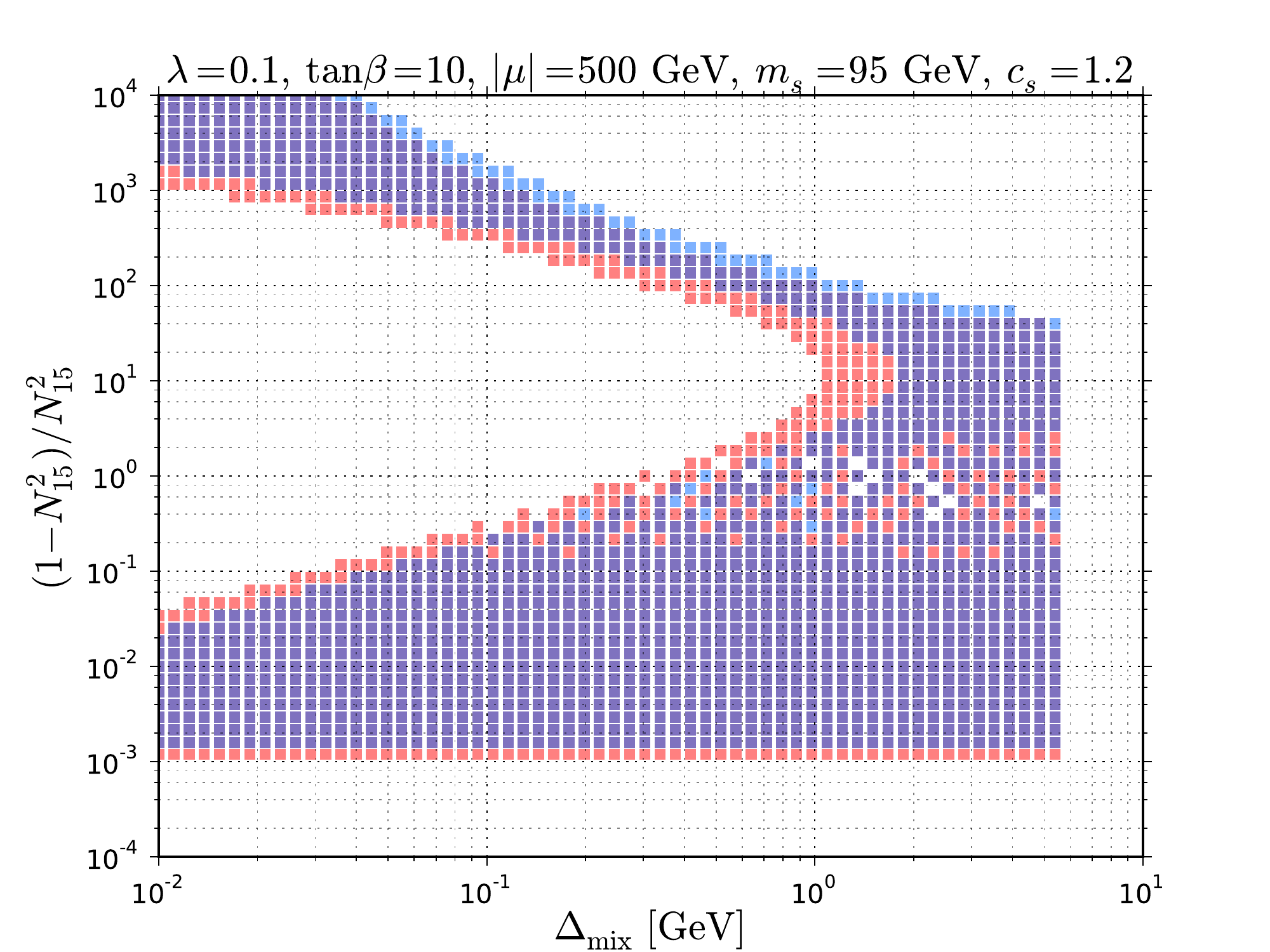}
\caption{Regions of the plane ($\Delta_{\rm mix}$, $(1-N_{15}^2)/N_{15}^2$) 
with $\sigma_{\rm SI}$ smaller than the neutrino background~\cite{NeutrinoB} 
for $m_{\chi}\mu>0$ (red) and $m_{\chi}\mu<0$ (blue), while keeping 
$|\kappa|\leq 0.6$. Upper (lower) plots correspond to $m_s=70$ (95) GeV 
whereas the left (right) to $c_s$ smaller (larger) than 1.
}
\label{fig:bs_fhs_N15_delta}
\end{figure}
\begin{equation}
\label{bs_fhs_mix-Higgsino}
\frac{\gamma+\mathcal{A}_s}{1-\gamma\mathcal{A}_s}\approx
{\rm sgn}(\mu)|N_{15}|\sqrt{2(1-{\rm sgn}\left(m_{\chi}\mu\right)\sin2\beta)}
\,.
\end{equation}
For specific values of $c_s$ and $m_s$ (chosen in our example) the l.h.s.\ 
of the above equation is proportional to $\gamma$ 
(with a negative coefficient\footnote{
It can be easily seen if we notice that $\mathcal{A}_s=-\gamma\;{\rm const}$, 
where ${\rm const}=\frac{1+c_s}{1+c_h}\big(\frac{m_h}{m_s}\big)^2>1$. Moreover 
$|\gamma\mathcal{A}_s|\ll 1$ and hence the denominator in the l.h.s.\ 
of~\eqref{bs_fhs_mix-Higgsino} is roughly 1.
})  
and thus to $\sqrt{\Delta_{\rm mix}}$ (see~\eqref{Delta_gamma})~-- 
this explains why there is a correlation between $\Delta_{\rm mix}$ and 
$|N_{15}|$. To understand why for $c_s>1$ $(c_s<1)$ there are (no) points which 
fulfill~\eqref{bs_fhs_mix-Higgsino} we should notice (see eqs~\eqref{Mhs}, 
\eqref{MHs}) that for $\tan\beta\gg1$ we have 
${\rm sgn}(1-c_s)={\rm sgn}(\Lambda\gamma)={\rm sgn}(\mu\gamma)$~-- 
the second equality holds because a partial cancellation between the two 
terms in $M_{h\hat s}^2$ is needed.\footnote{
This happens in our example in Fig.~\ref{fig:bs_fhs_N15_delta} because 
for $|\mu|=500$~GeV and $\lambda=0.1$ we have 
$|M_{h\hat s}^2|\sim\mathcal O(100\;{\rm GeV})$, which is of order $m_h$ 
and $m_s$. The situation for smaller $|\mu|$ is not much different.
} 
This is exactly what we wanted to show: for $c_s<1$ the l.h.s.\ 
of~\eqref{bs_fhs_mix-Higgsino} has the sign equal to $-{\rm sgn}(\mu)$
thus the equality cannot hold (and inversely for $c_s>1$). It can be shown 
(using relations~\eqref{Nj3Nj5} and \eqref{Nj4Nj5}),
that the above conclusions hold also for some part of a highly mixed LSP 
parameter space when $|\kappa/\lambda|$ is smaller than
$|\frac{N_{13}}{N_{15}}\frac{N_{14}}{N_{15}}|$ i.e.\ with unsuppressed $|\eta|$ 
in eq.~\eqref{bs_fhs_mix}. For a singlino-dominated LSP we can always choose the
sign and value of $\kappa$ to fulfill relation~\eqref{bs_fhs_mix}.

Let us finally comment on the fact that for large $\tan\beta$ the $H$ 
exchange might be relevant if $H$ is light enough. The presence of relatively 
light $H$ usually results in stronger constraints on the parameter space, 
especially for large values of $\lambda$. This is because in this region 
of the parameter space $|M_{\hat{H}\hat{s}}^2|$ is well approximated by 
$\lambda v \mu \tan\beta$ so it is typically larger than the diagonal 
entries of the Higgs mass matrix, unless $\lambda$ is small. As a result, 
large values of $\lambda$ lead to tachyons, or at least the mixing effects 
that are too large to accommodate the LEP and/or LHC Higgs data.

\section{Summary}
\label{sec:sum}

We have investigated blind spots for spin-independent scattering cross-section 
for the Higgsino-singlino LSP in the NMSSM. If mixing between the 
(SM-like) Higgs scalar and other scalars is negligible, a blind spot can 
occur only if the ratio $m_{\chi}/\mu$ is positive and has value close to 
$\sin2\beta$. Then, blind spots exist only for singlino-dominated LSPs 
(unless $\tan\beta$ is very close to 1) with the amount of the Higgsino 
component determined by $\tan\beta$. This changes a lot when mixing with 
the singlet scalar is taken into account.

If the singlet-dominated scalar is heavier than the Higgs scalar, the 
Higgs-singlet mixing has to be quite small to avoid large negative correction 
to the Higgs scalar mass. But even for such small mixing new classes of blind 
spots appear. Blind spots for Higgsino-dominated LSPs become possible and 
the ratio $m_{\chi}/\mu$ may be also negative. The LSP composition is no 
longer so strongly related to $\tan\beta$, especially for smaller values of 
$\lambda$. However, in most cases the LSP must be highly dominated either 
by Higgsino or by singlino. A blind spot for a highly mixed 
Higgsino-singlino LSP is possible only for small values of $\lambda$ and 
$\tan\beta$ and positive $m_{\chi}/\mu$. In addition, in the most often 
explored part of NMSSM parameter space with large (but perturbative) 
$\lambda$ and small $\tan\beta$, a blind spot for a singlino-dominated LSP 
can occur only if $m_{\chi}\mu>0$ and eq.~\eqref{bs_fh_0} is approximately 
satisfied.

If the singlet-dominated scalar is lighter than the Higgs scalar, 
large Higgs-singlet mixing is welcome because the contribution from such 
mixing to the Higgs scalar mass is positive. For small $\tan\beta$, the LEP 
and LHC constraints allow for sizable mixing leading to the correction to 
the Higgs scalar mass $\Delta_{\rm mix}\sim5$ GeV for the singlet mass in the 
range of about 85$\div$105 GeV. For such big $\Delta_{\rm mix}$, a blind spot 
for large $\lambda$ and $\tan\beta\sim2$ may occur also for highly mixed 
Higgsino-singlino LSP if $m_{\chi}\mu>0$, which would not be possible otherwise. 
It should be noted, however, that not always large $\Delta_{\rm mix}$ is 
beneficial for a blind spot occurrence. For example, for an LSP strongly 
dominated by the Higgsino a blind spot may occur only if $\Delta_{\rm mix}$ 
is small.

For light singlet scalar and big $\Delta_{\rm mix}$ the region of moderate 
and large $\tan\beta$ is also interesting. In such a case the singlet coupling 
to bottom quarks may be significantly different than the one to gauge bosons. 
If the $sb\bar{b}$ coupling is suppressed, relatively large $\Delta_{\rm mix}$ 
is allowed by LEP also for $m_s<85$ GeV. We found that for suppressed  
$sb\bar{b}$ coupling a blind spot may occur only for a singlino-dominated LSP. 
On the other hand, if the $sb\bar{b}$ coupling is enhanced a blind spot can 
exist for any composition of the LSP and for both signs of $m_{\chi}\mu$.

For large $\tan\beta$ one more class of blind spots may exist if the heavier 
scalar doublet $H$ is light enough to mediate the LSP-nucleon interaction in 
a substantial way and the singlet-dominated scalar is rather heavy. 
In such a case, positive $m_{\chi}\mu$ is again preferred, 
allowing for blind spots for the LSP composition much less restricted 
than in the case with very heavy $H$. 
If the Higgs-singlet mixing is present, $m_{\chi}\mu<0$ is also possible  
but in this case the influence of a relatively light $H$ on possible 
blind spots is quite marginal. In addition, smaller values of $m_H$ 
result in stronger upper bounds on the coupling $\lambda$.

There are several avenues for future studies where the results obtained 
in this paper can be used. For instance, it will be crucial to
investigate how one can probe neutralino LSP with SI scattering cross-section 
below the neutrino background. 
Some possible ways to constrain blind spots may be to 
use the direct and indirect detection experiments sensitive to the SD 
cross-sections or dedicated collider searches which in the context 
of MSSM turn out to be complementary to direct dark matter searches, see 
e.g.~\cite{underabundant,Barducci,Cao:2015efs} for some recent work on this 
topic. Some studies of the LHC sensitivity to Higgsino-singlino sector has 
already been done \cite{EllwangerHiggsinoSinglino} but more effort in this 
direction is welcome. It will be also interesting to investigate whether 
the blind spots identified in this paper can exist in more constrained 
versions of NMSSM and in which scenarios it is possible to explain the 
observed abundance of dark matter assuming thermal history of the Universe. 
We plan to investigate these issues in the future.

\section*{Acknowledgments}

This work was partially supported by Polish National Science Centre 
under research grants DEC-2012/05/B/ST2/02597, DEC-2014/15/B/ST2/02157
and DEC-2012/04/A/ST2/00099. MB acknowledges support from the Polish 
Ministry of Science and Higher Education (decision no.\ 1266/MOB/IV/2015/0).
MB thanks the Galileo Galilei Institute for Theoretical Physics and INFN 
for hospitality and partial support during the completion of this work.

\section*{Useful formulae}
%
The parameter $\eta$ defined in \eqref{eta_def} may be 
expressed in terms of other parameters of the NMSSM model. 
With the help of eqs.~\eqref{Nj3Nj5} and \eqref{Nj4Nj5} one can write 
it as the following function of three dimensionless ratios: 
$(\lambda v)/\mu$, $\kappa/\lambda$ and $m_{\chi}/\mu$:
\begin{equation}
\label{eta_lambda/mu}
\eta
=
\frac{\frac{\lambda v}{\mu}\,
\left(1-\left(\frac{m_{\chi}}{\mu}\right)^2\right)
\left(\frac{m_{\chi}}{\mu}-\sin2\beta\right)}
{
\left(\frac{\lambda v}{\mu}\right)^2
\left[\left(1+\left(\frac{m_{\chi}}{\mu}\right)^2\right)\frac{\sin2\beta}{2}-\frac{m_{\chi}}{\mu}\right]
-\frac{\kappa}{\lambda}\left(1-\left(\frac{m_{\chi}}{\mu}\right)^2\right)^2
}\,.
\end{equation}
Equation \eqref{Higgsino/singlino} may be used to eliminate 
$(\lambda v)/\mu$ in favor of the ratio 
$(1-N_{15}^2)/N_{15}^2$ characterizing the LSP composition. 
Then, one obtains another expression for $\eta$:
\begin{equation}
\label{eta-N15}
\eta
=
\frac
{\sgn(\mu)\left(\frac{m_\chi}{\mu}-\sin2\beta\right)}
{
\frac12\sqrt{\frac{1-N_{15}^2}{N_{15}^2}}
\frac{\left(1+\left(\frac{m_\chi}{\mu}\right)^2\right)\sin2\beta-2\frac{m_\chi}{\mu}}
{\sqrt{1+\left(\frac{m_\chi}{\mu}\right)^2-2\frac{m_\chi}{\mu}\sin2\beta}}
-\frac{\kappa}{\lambda}\,\sqrt{\frac{N_{15}^2}{1-N_{15}^2}}
\sqrt{1+\left(\frac{m_\chi}{\mu}\right)^2-2\frac{m_\chi}{\mu}\sin2\beta}
}\,.
\end{equation}
It will be helpful to consider a few limits of this parameter.
Let us start with the situation when one of the terms in the denominator 
dominates over the other one.
The first (second) term in the denominator may be neglected if 
$|\frac{\kappa}{\lambda}|$ is much bigger (smaller) than 
$\frac{1-N_{15}^2}{2N_{15}^2}
\left|
\frac{\left(m_\chi/\mu+\mu/m_\chi\right)\sin2\beta-2}
{\left(m_\chi/\mu+\mu/m_\chi\right)-2\sin2\beta}
\right|$.
The second factor in the last expression is always smaller than 1 and 
approaches 1 in the limit $|m_\chi/\mu|\to1$ i.e.\ for a strongly 
Higgsino-dominated LSP. It may be very small if $m_\chi\mu>0$ and 
$\sin2\beta\approx2/\left(m_\chi/\mu+\mu/m_\chi\right)$.

If $\left|\frac{\kappa}{\lambda}\right|\gg
\frac{1-N_{15}^2}{2N_{15}^2}
\left|
\frac{\left(m_\chi/\mu+\mu/m_\chi\right)\sin2\beta-2}
{\left(m_\chi/\mu+\mu/m_\chi\right)-2\sin2\beta}
\right|
>\frac{1-N_{15}^2}{2N_{15}^2}$
i.e.\ we are considering a singlino-dominated LSP and/or $|\kappa|$ much 
bigger than $\lambda$ (for a not strongly Higgsino-dominated LSP),  
the parameter $\eta$ is approximately given by:
\begin{equation}
\label{eta-big_kappa}
\eta
\approx
-{\rm sgn}(\mu)\frac{\lambda}{\kappa}\sqrt{\frac{1-N_{15}^2}{N_{15}^2}}
\frac{\frac{m_{\chi}}{\mu}-\sin2\beta}
{\sqrt{1+\left(\frac{m_{\chi}}{\mu}\right)^2-2\frac{m_{\chi}}{\mu}\sin2\beta}}\,.
\end{equation}
If $\left|\frac{\kappa}{\lambda}\right|\ll
\frac{1-N_{15}^2}{2N_{15}^2}
\left|
\frac{\left(m_\chi/\mu+\mu/m_\chi\right)\sin2\beta-2}
{\left(m_\chi/\mu+\mu/m_\chi\right)-2\sin2\beta}
\right|$ 
i.e.\ for a Higgsino-dominated LSP and/or $|\kappa|$ is much smaller than 
$\lambda$ (for a not strongly singlino-dominated LSP) we get:
\begin{equation}
\label{eta-small_kappa}
\eta
\approx
{\rm sgn}(\mu)\sqrt{\frac{N_{15}^2}{1-N_{15}^2}}
\frac{2\left(\frac{m_{\chi}}{\mu}-\sin2\beta\right)
\sqrt{1+\left(\frac{m_{\chi}}{\mu}\right)^2-2\frac{m_{\chi}}{\mu}\sin2\beta}}
{\left(1+\left(\frac{m_{\chi}}{\mu}\right)^2\right)\sin2\beta-2\frac{m_{\chi}}{\mu}}\,.
\end{equation}
In the case of a strongly Higgsino-dominated LSP, 
using $N_{15}^2\ll 1$ and $m_\chi^2\approx\mu^2$,
the above equality may be further approximated as:
\begin{equation}
\label{eta-Higgsino}
\eta
\approx
-{\rm sgn}(\mu)|N_{15}|\sqrt{2(1-{\rm sgn}\left(m_{\chi}\mu\right)\sin2\beta)}
\,.
\end{equation}

Above there are several forms and limits of the parameter $\eta$ 
defined in eq.~\eqref{eta_def}. With the help of 
eqs.~\eqref{Nj3Nj5}-\eqref{Higgsino/singlino} one may 
rewrite also the whole amplitude \eqref{fN-B} as:
\begin{align}
f^{(N)}
\approx&
\frac{\lambda}{\sqrt{2}}\frac{\alpha_{hNN}}{m_h^2}\tilde{S}_{h\hat{h}}
\left.
\sqrt{\frac{N_{15}^2(1-N_{15}^2)}
{1+\left(\frac{m_{\chi}}{\mu}\right)^2-2\frac{m_{\chi}}{\mu}\sin2\beta}}
\right\{
\mathcal{B}_{\hat{h}} \left[\frac{m_{\chi}}{\mu}-\sin2\beta\right]
+\mathcal{B}_{\hat{H}} \cos2\beta
\nn
&\qquad\left.
+\mathcal{B}_{\hat{s}}
\left[
\frac{\frac{\lambda v}{\mu}}{1-\left(\frac{m_{\chi}}{\mu}\right)^2}
\left(\left(1+\left(\frac{m_{\chi}}{\mu}\right)^2\right)
\frac{\sin2\beta}{2}-\frac{m_{\chi}}{\mu}\right)
-\frac{\kappa}{\lambda}\,
\frac{1-\left(\frac{m_{\chi}}{\mu}\right)^2}{\frac{\lambda v}{\mu}}
\right]
\right\}.
\label{fNmixed}
\end{align}
This formula is not very convenient in the case of a strongly 
Higgsino-dominated LSP. In such limit $N_{15}\to0$ but one of the terms 
in the curly bracket diverges as $m_{\chi}^2\to\mu^2$. Thus, for a strongly 
Higgsino-dominated LSP it is better to rewrite eq.~\eqref{fNmixed} 
in the following form:
\begin{align}
f^{(N)}
&\approx
\frac{\lambda}{\sqrt{2}}\frac{\alpha_{hNN}}{m_h^2}
\frac{\tilde{S}_{h\hat{h}}\left(1-N_{15}^2\right)}
{1+\left(\frac{m_{\chi}}{\mu}\right)^2-2\frac{m_{\chi}}{\mu}\sin2\beta}
\left\{
\mathcal{B}_{\hat{h}}
\frac{1-\left(\frac{m_{\chi}}{\mu}\right)^2}{\frac{\lambda v}{\mu}}
\left[\frac{m_{\chi}}{\mu}-\sin2\beta\right]
\right.
\nn
&\left.
+\mathcal{B}_{\hat{H}}
\frac{1-\left(\frac{m_{\chi}}{\mu}\right)^2}{\frac{\lambda v}{\mu}}
\cos2\beta
+\mathcal{B}_{\hat{s}}
\left[
\left(1+\left(\frac{m_{\chi}}{\mu}\right)^2\right)
\frac{\sin2\beta}{2}-\frac{m_{\chi}}{\mu}
-\frac{\kappa}{\lambda}
\left(\frac{1-\left(\frac{m_{\chi}}{\mu}\right)^2}{\frac{\lambda v}{\mu}}
\right)^2
\right]
\right\}.
\end{align}

\vspace{1ex}
\noindent
{\bf Comments on the spin-dependent scattering cross-section}
\vspace{1ex}

\noindent
The only contribution at the tree-level to the spin-dependent scattering 
cross-section in our case comes from the $t$-channel $Z$ exchange, so depends 
only on the Higgsino contribution to the LSP and reads:
\begin{equation}
\label{eq:sigSD}
\sigma_{\rm SD}^{(N)}=C^{(N)}\cdot 10^{-38}\;{\rm cm^2}\;\,(N_{13}^2-N_{14}^2)^2\,,
\end{equation}
where $C^{(p)}\approx 4$, $C^{(n)}\approx 3.1$~\cite{SD}.
Combining eqs.~\eqref{Nj3Nj5}, \eqref{Nj4Nj5} and \eqref{Higgsino/singlino} 
we can write:  
\begin{equation}
\label{eq:N13N14diff}
N_{13}^2-N_{14}^2=\frac{\left[1-\left(m_\chi/\mu\right)^2\right](1-N_{15}^2)\cos2\beta}{1+\left(m_\chi/\mu\right)^2-2\left(m_\chi/\mu\right)\sin2\beta}\,.
\end{equation}
We can see immediately that the cross-section disappear in the limit 
of $\tan\beta=1$ or a pure singlino/Higgsino LSP. 
Using eq.~\eqref{Higgsino/singlino} we may rewrite the last formula 
in the form
\begin{equation}
\label{eq:N13N14diff_lambda}
N_{13}^2-N_{14}^2=\frac{\left[1-\left(m_\chi/\mu\right)^2\right]
\cos2\beta}{1+\left(m_\chi/\mu\right)^2-2\left(m_\chi/\mu\right)\sin2\beta
+\left[1-\left(m_\chi/\mu\right)^2\right]^2
\left({\mu}/{\lambda v}\right)^2}\,
\end{equation}
showing the explicit dependence of the LSP-$Z$ coupling on $\lambda$ 
(there is also an implicit dependence via the LSP mass $m_\chi$).



\begin{thebibliography}{99}

\bibitem{Atlas_discovery}
  G.~Aad {\it et al.}  [ATLAS Collaboration],
  Phys.\ Lett.\ B {\bf 716} (2012) 1
  [arXiv:1207.7214 [hep-ex]].

\bibitem{CMS_discovery}
  S.~Chatrchyan {\it et al.}  [CMS Collaboration],
  Phys.\ Lett.\ B {\bf 716} (2012) 30
  [arXiv:1207.7235 [hep-ex]].

\bibitem{LUX}
  D.~S.~Akerib {\it et al.} [LUX Collaboration],
  Phys.\ Rev.\ Lett.\  {\bf 112} (2014) 091303
  [arXiv:1310.8214 [astro-ph.CO]].
  
\bibitem{XENON1T}
  E.~Aprile {\it et al.} [XENON Collaboration],
  [arXiv:1512.07501 [physics.ins-det]].

\bibitem{LZ}
  D.~C.~Malling {\it et al.},
  arXiv:1110.0103 [astro-ph.IM].

\bibitem{NeutrinoB}
  J.~Billard, L.~Strigari and E.~Figueroa-Feliciano,
  Phys.\ Rev.\ D {\bf 89} (2014) 2,  023524
  [arXiv:1307.5458 [hep-ph]].
  
\bibitem{Planck}
  P.~A.~R.~Ade {\it et al.}  [Planck Collaboration],
  arXiv:1502.01589 [astro-ph.CO].
  
\bibitem{Hall}
  C.~Cheung, L.~J.~Hall, D.~Pinner and J.~T.~Ruderman,
  JHEP {\bf 1305} (2013) 100
  [arXiv:1211.4873 [hep-ph]].
  
  \bibitem{Wagner}
  P.~Huang and C.~E.~M.~Wagner,
  Phys.\ Rev.\ D {\bf 90} (2014) 1,  015018
  [arXiv:1404.0392 [hep-ph]].
  
\bibitem{reviewEllwanger}
  U.~Ellwanger, C.~Hugonie and A.~M.~Teixeira,
  Phys.\ Rept.\  {\bf 496} (2010) 1
  [arXiv:0910.1785 [hep-ph]].


\bibitem{Greene:1986th}
  B.~R.~Greene and P.~J.~Miron,
  Phys.\ Lett.\ B {\bf 168} (1986) 226.

\bibitem{Flores:1990bt}
  R.~Flores, K.~A.~Olive and D.~Thomas,
  Phys.\ Lett.\ B {\bf 245} (1990) 509.

 \bibitem{Belanger:2005kh}
  G.~Belanger, F.~Boudjema, C.~Hugonie, A.~Pukhov and A.~Semenov,
  JCAP {\bf 0509} (2005) 001
  [hep-ph/0505142].

  
\bibitem{NMSSM_DD}
  D.~G.~Cerdeno, C.~Hugonie, D.~E.~Lopez-Fogliani, C.~Munoz and A.~M.~Teixeira,
  JHEP {\bf 0412} (2004) 048
  [hep-ph/0408102];
  D.~G.~Cerdeno, E.~Gabrielli, D.~E.~Lopez-Fogliani, C.~Munoz and A.~M.~Teixeira,
  JCAP {\bf 0706} (2007) 008
  [hep-ph/0701271 ].
  
\bibitem{NMSSM_DD_Barger}
  V.~Barger, P.~Langacker, I.~Lewis, M.~McCaskey, G.~Shaughnessy and B.~Yencho,
  Phys.\ Rev.\ D {\bf 75} (2007) 115002
  [hep-ph/0702036 [HEP-PH]].
  
\bibitem{Das:2010ww}
  D.~Das and U.~Ellwanger,
  JHEP {\bf 1009} (2010) 085
  [arXiv:1007.1151 [hep-ph]].

\bibitem{Kozaczuk:2013spa}
  J.~Kozaczuk and S.~Profumo,
  Phys.\ Rev.\ D {\bf 89} (2014) 9,  095012
  [arXiv:1308.5705 [hep-ph]].
  
\bibitem{Cao:2013mqa}
  J.~Cao, C.~Han, L.~Wu, P.~Wu and J.~M.~Yang,
  JHEP {\bf 1405} (2014) 056
  [arXiv:1311.0678 [hep-ph]].

  
  
\bibitem{Han:2014nba}
  T.~Han, Z.~Liu and S.~Su,
  JHEP {\bf 1408} (2014) 093
  [arXiv:1406.1181 [hep-ph]].
  
  \bibitem{Enberg:2015qwa}
  R.~Enberg, S.~Munir, C.~P\'{e}rez de los Heros and D.~Werder,
  arXiv:1506.05714 [hep-ph].

  \bibitem{mixedDM_Cheung}
  C.~Cheung and D.~Sanford,
  JCAP {\bf 1402} (2014) 011
  [arXiv:1311.5896 [hep-ph]].
  
  \bibitem{mixedDM_Calibbi}
  L.~Calibbi, A.~Mariotti and P.~Tziveloglou,
  arXiv:1505.03867 [hep-ph].
        
\bibitem{BaOlPo}
M.~Badziak, M.~Olechowski and S.~Pokorski,
  JHEP {\bf 1306} (2013) 043
  [arXiv:1304.5437 [hep-ph]].
  
\bibitem{LEP2}
  S.~Schael {\it et al.}  [ALEPH and DELPHI and L3 and OPAL and LEP Working Group for Higgs Boson Searches Collaborations],
  Eur.\ Phys.\ J.\ C {\bf 47} (2006) 547
  [hep-ex/0602042];
  [LEP Higgs Working Group for Higgs boson searches Collaboration],
  hep-ex/0107034.

\bibitem{NMSSM_DD_EllwangerHugonie}
  U.~Ellwanger and C.~Hugonie,
  JHEP {\bf 1408} (2014) 046
  [arXiv:1405.6647 [hep-ph]].

\bibitem{underabundant}
  M.~Badziak, A.~Delgado, M.~Olechowski, S.~Pokorski and K.~Sakurai,
  JHEP {\bf 1511} (2015) 053
  [arXiv:1506.07177 [hep-ph]].
  
\bibitem{JuKaGr}
  G.~Jungman, M.~Kamionkowski and K.~Griest,
  Phys.\ Rept.\  {\bf 267} (1996) 195
  [hep-ph/9506380].

\bibitem{NTools1}
  U.~Ellwanger, J.~F.~Gunion and C.~Hugonie,
  JHEP {\bf 0502} (2005) 066
  [hep-ph/0406215].

\bibitem{NTools2}
  U.~Ellwanger and C.~Hugonie,
  Comput.\ Phys.\ Commun.\  {\bf 175} (2006) 290
  [hep-ph/0508022].
  
\bibitem{Belanger:2013oya}
  G.~Belanger, F.~Boudjema, A.~Pukhov and A.~Semenov,
  Comput.\ Phys.\ Commun.\  {\bf 185} (2014) 960
  [arXiv:1305.0237 [hep-ph]].

\bibitem{XENON100}
  E.~Aprile {\it et al.} [XENON100 Collaboration],
  Phys.\ Rev.\ Lett.\  {\bf 111} (2013) 2,  021301
  [arXiv:1301.6620 [astro-ph.CO]].

\bibitem{IceCubeNEW}
  M.~G.~Aartsen {\it et al.} [IceCube Collaboration],
  arXiv:1601.00653 [hep-ph].
     
\bibitem{lambdaSUSY}
  R.~Barbieri, L.~J.~Hall, Y.~Nomura and V.~S.~Rychkov,
  Phys.\ Rev.\ D {\bf 75} (2007) 035007
  [hep-ph/0607332];
    L.~J.~Hall, D.~Pinner and J.~T.~Ruderman,
  JHEP {\bf 1204} (2012) 131
  [arXiv:1112.2703 [hep-ph]].
  
\bibitem{Hisano}
  J.~Hisano, K.~Ishiwata, N.~Nagata and T.~Takesako,
  JHEP {\bf 1107} (2011) 005
  [arXiv:1104.0228 [hep-ph]];
  J.~Hisano, K.~Ishiwata and N.~Nagata,
  JHEP {\bf 1506} (2015) 097
  [arXiv:1504.00915 [hep-ph]];
R.~J.~Hill and M.~P.~Solon,
  Phys.\ Lett.\ B {\bf 707} (2012) 539
  [arXiv:1111.0016 [hep-ph]].
  
\bibitem{ATLAStautau}
  G.~Aad {\it et al.} [ATLAS Collaboration],
  JHEP {\bf 1411} (2014) 056
  [arXiv:1409.6064 [hep-ex]].
  
\bibitem{CMStautau}
  V.~Khachatryan {\it et al.} [CMS Collaboration],
  JHEP {\bf 1410} (2014) 160
  [arXiv:1408.3316 [hep-ex]].
  
\bibitem{Cao:2015loa}
  J.~Cao, L.~Shang, P.~Wu, J.~M.~Yang and Y.~Zhang,
  JHEP {\bf 1510} (2015) 030
  [arXiv:1506.06471 [hep-ph]].
  
\bibitem{Barducci}
  D.~Barducci, A.~Belyaev, A.~K.~M.~Bharucha, W.~Porod and V.~Sanz,
  JHEP {\bf 1507} (2015) 066
  [arXiv:1504.02472 [hep-ph]].
  
  \bibitem{Cao:2015efs}
  J.~Cao, Y.~He, L.~Shang, W.~Su and Y.~Zhang,
  arXiv:1511.05386 [hep-ph].

\bibitem{EllwangerHiggsinoSinglino}
  U.~Ellwanger,
  JHEP {\bf 1311} (2013) 108
  [arXiv:1309.1665 [hep-ph]].

\bibitem{SD}
  G.~Chalons, M.~J.~Dolan and C.~McCabe,
  JCAP {\bf 1302} (2013) 016
  [arXiv:1211.5154 [hep-ph]];
  G.~Belanger, F.~Boudjema, A.~Pukhov and A.~Semenov,
  Comput.\ Phys.\ Commun.\  {\bf 180} (2009) 747
  [arXiv:0803.2360 [hep-ph]].
    
\end{thebibliography}
\end{document}